\documentclass[Journal]{IEEEtran}

\usepackage{amssymb,color}
\usepackage{amsmath}
\usepackage{bbm}
\usepackage{graphicx}
\usepackage{epstopdf}
\usepackage{amsthm,amsfonts}
\usepackage{subcaption}
\usepackage{setspace}
\usepackage{lipsum}
\usepackage{multicol}

\usepackage{accents}

\usepackage{algorithm}
\usepackage{algorithmic}
\usepackage{psfrag}
\usepackage{cite}
\usepackage{pifont}
\usepackage[font={small}]{caption}

\allowdisplaybreaks

\makeatletter
\newcommand{\oset}[3][0ex]{%
  \mathrel{\mathop{#3}\limits^{
    \vbox to#1{\kern-2\ex@
    \hbox{$\scriptstyle#2$}\vss}}}}
\makeatother

\begin{document}
\title{Downlink Non-Orthogonal Multiple Access (NOMA) in Poisson Networks} 

\author{
\IEEEauthorblockN{\large  Konpal Shaukat Ali$^*$,~\IEEEmembership{Student Member,~IEEE,} Martin Haenggi$^{\dagger}$,~\IEEEmembership{Fellow,~IEEE,} Hesham ElSawy$^{\star}$,~\IEEEmembership{Senior Member,~IEEE,} Anas Chaaban$^{\ddagger}$,~\IEEEmembership{Senior Member,~IEEE,} and Mohamed-Slim Alouini$^*$,~\IEEEmembership{Fellow,~IEEE}}

\thanks{$^*$The authors are with the Computer, Electrical, and Mathematical Sciences and Engineering (CEMSE) Divison, King Abdullah University of Science and Technology (KAUST), Thuwal, Makkah Province, Saudi Arabia. (Email: \{konpal.ali, slim.alouini\}@kaust.edu.sa)

$^{\dagger}$ The author is with the Department of Electrical Engineering, University of Notre Dame, USA. (Email:mhaenggi@nd.edu)

$^{\star}$ The author is with the Department of Electrical Engineering, King Fahd University of Petroleum and Minerals (KFUPM), Dhahran, Saudi Arabia. (Email: hesham.elsawy@kfupm.edu.sa)

$^{\ddagger}$ The author is with the School of Engineering, University of British Columbia, Kelowna, BC  Canada V1V 1V7. (Email: anas.chaaban@ubc.ca)

{Part of this work was presented at the IEEE International Conference on Communications (ICC'18) \cite{myNOMA_icc}.}

M. Haenggi gratefully acknowledges the support of the U.S.~National Science Foundation through grant CCF 1525904.
 }}

\maketitle

\begin{abstract} 

A network model is considered where Poisson distributed base stations transmit to $N$ power-domain non-orthogonal multiple access (NOMA) users (UEs) each {that employ successive interference cancellation (SIC) for decoding}. We propose three models for the clustering of NOMA UEs and consider two different ordering techniques for the NOMA UEs: mean signal power-based and instantaneous signal-to-intercell-interference-and-noise-ratio-based. For each technique, we present a signal-to-interference-and-noise ratio analysis for the coverage of the typical UE. We plot the rate region for the two-user case and show that neither ordering technique is consistently superior to the other. We propose two efficient algorithms for finding a feasible resource allocation that maximize the cell sum rate $\mathcal{R}_{\rm tot}$, for general $N$, constrained to: 1) a minimum throughput $\mathcal{T}$ for each UE, 2) identical throughput for all UEs. We show the existence of: 1) an optimum $N$ that maximizes the constrained $\mathcal{R}_{\rm tot}$ given a set of network parameters, 2) a critical SIC level necessary for NOMA to outperform orthogonal multiple access. The results highlight the importance in choosing the network parameters $N$, the constraints, and the ordering technique to balance the $\mathcal{R}_{\rm tot}$ and fairness requirements. We also show that interference-aware UE clustering can significantly improve performance.


\end{abstract}


\section{Introduction}

The available spectrum is a scarce resource, and many new technologies to be incorporated into 5G aim at reusing the spectrum more efficiently to improve data rates and fairness. Traditionally, temporal, spectral, or spatial\footnote{Spatial separation of UEs with MIMO can be used with either OMA or NOMA.} orthogonalization techniques, referred to as orthogonal multiple access (OMA), are used to avoid interference among users (UEs) in a cell. They allow only one UE per time-frequency resource block in a cell. A promising candidate for more efficient spectrum reuse in 5G is non-orthogonal multiple access (NOMA), which allows multiple UEs to share the same time-frequency resource block. {The set of UEs being served by a base station (BS) via NOMA is referred to as the UE cluster. A UE cluster is served} by having messages multiplexed either in the power domain or in the code domain. NOMA is therefore a special case of superposition coding \cite{N12_9}. Decoding techniques using successive interference cancellation (SIC) \cite{book_fundWirelessComm} for multiple-access channels have been studied from an information-theoretic perspective for several decades \cite{N12_16}, and they were implemented on a software radio platform in \cite{sup1}. The focus of our work is on NOMA where messages are superposed in the power domain. This form of NOMA allows multiple UEs to transmit/receive messages in the same time-frequency resource block by transmitting them at different power levels. SIC techniques are then used for decoding.

With NOMA come a number of challenges, including:
\begin{enumerate}
\item Determining the size of the UE cluster, i.e., the number of UEs to be served by a BS.
\item Determining which UEs to include in a cluster, referred to as UE clustering.
\item Ordering UEs within a cluster according to some measure of link quality.
\item The objective of the cluster $-$ prioritizing individual UE performance, total cluster performance, or a middle ground between the two.
\item Resource allocation (RA) for the UEs in a cluster.
\end{enumerate}
Promising results for NOMA as an efficient spectrum reuse technique have been shown \cite{N1,N8}. In \cite{N9,N13} power allocation (PA) schemes are investigated for universal fairness by achieving identical rates for NOMA UEs. The idea of cooperative NOMA is investigated in \cite{N7,N4}. Most NOMA works order UEs based on either their distance from the transmitting BS \cite{N2,N13,N4,N6} or on the quality of the transmission channel \cite{N5,N15,N9,N7,N8}. A number of works in the literature focus on RA \cite{N15,N9,N13,N14,N15_19}. RA schemes for maximizing rates with constraints {often focus on} small NOMA clusters such as the two-user case \cite{N13,N14,N15_19}, though works such as \cite{N15,N9} consider a general number of UEs in a NOMA cluster. 

{The works in \cite{N1,N8,N9,N13,N7,N4,N15,N14,N15_19} consider NOMA in a single cell and therefore do not account for intercell interference, {denoted by $I^{\rm\o}$}, which has a drastic negative impact on the NOMA performance as shown in \cite{my_nomaMag}. Stochastic geometry has succeeded in providing a unified mathematical paradigm to model large cellular networks and characterize their operation while accounting for intercell interference \cite{MH_Book2,3B1,h_tut,di_renzo}. Using stochastic geometry-based modeling, a large uplink NOMA network is studied in \cite{N6,N18}, a large downlink NOMA network in \cite{N16,N18}, and a qualitative study on NOMA in large networks is carried out in \cite{my_nomaMag}.} However, \cite{N16} does not take into account the SIC chain in the signal-to-interference-and-noise ratio (SINR) analysis, which overestimates coverage. {In \cite{N18}, two-user NOMA with fixed PA is studied. In the downlink, comparisons are made between random UE selection and selection such that the weaker UE's channel quality is below a threshold and the stronger UE's is greater than a second threshold.} {It ought to be mentioned that fixed RA does not allow the system to meet a defined cluster objective and makes {a comparison} with other schemes such as OMA unfair.}

In this work we analytically study a large multi-cell downlink NOMA system that takes into account intercell interference and intracell interference, henceforth called \emph{intraference} and {denoted by $I^\circ$}, error propagation in the SIC chain, and the effects of imperfect SIC for a general number of UEs served by each BS {(i.e., a general cluster size). We discuss all of the challenges enumerated above.} Our goal is to analyze the performance of such a large network setup using stochastic geometry. {We introduce and study three different models to show the impact of location-based selection of NOMA UEs in a cluster, i.e., UE clustering, on performance.} {We analyze and compare the network performance using two ordering techniques, namely mean signal power- (MSP-) based ordering, which is equivalent to distance-based ordering, and instantaneous signal-to-intercell-interference-and-noise-ratio- (IS$\oset[-.1ex]{_{\o}}{\raise.01ex\hbox{I}}$NR-) based ordering. {In this context, the rate region for the two-user case is studied for both ordering techniques.} {To the best of our knowledge, an analytical work that compares both ordering techniques does not exist.} We consider two main objectives: 1) maximizing the cell sum rate defined as the sum of the throughput of all the UEs in a NOMA cluster of the cell, subject to a threshold minimum throughput (TMT) constraint {of $\mathcal{T}$} on the individual UEs 2) maximizing the cell sum rate when all UEs in a cluster have identical throughput, i.e., maximizing the symmetric throughput. Accordingly, {we formulate optimization problems and} propose {algorithms for} intercell interference-aware RA for both objectives.} {We show a significant reduction in the complexity of our proposed algorithms when compared to an exhaustive search.} OMA is used to benchmark the gains attained by NOMA. 



The contributions of this paper can be summarized as follows:
\begin{itemize}
\item {We propose three models {for the clustering of UEs (i.e., UE selection), which are governed by two important principles: First, a UE should be served by its closest (or strongest) BS; conversely, a BS chooses its NOMA UE from among the UEs in its Voronoi cell. Second, only UEs with good channel conditions (on average) should be served using NOMA (i.e., sharing resource blocks).} In contrast, using a standard Matern cluster process such as in \cite{N6} would lead to the unrealistic situation where UEs from another Voronoi cell may be part of a NOMA cluster.}
\item {From the rate region for the two-user case we show that contrary to the expected result UE ordering based on IS$\oset[-.1ex]{_{\o}}{\raise.01ex\hbox{I}}$NR, which takes into account information about not only the path loss but also fading, intercell interference and noise, is not always superior to MSP-based ordering. We discuss how RA and intraference impact this finding.}
\item We show that there exists an optimum NOMA cluster size that maximizes the constrained cell sum rate given the residual intraference (RI) factor $\beta$.
\item {We show the existence of a critical level of SIC $1-\beta$ that is necessary for NOMA to outperform OMA.}
\end{itemize}

The rest of the paper is organized as follows: Section II describes the system model. The SINR analysis and relevant statistics are in Section III. In Section IV the two optimization problems are formulated, the rate region for the two-user scenario is discussed, and algorithms for solving the problems are proposed. Section V presents the results, and Section VI concludes the paper. 

\textit{Notation:} Vectors are denoted using bold text, $\|\textbf{x} \|$ denotes the Euclidean norm of the vector $\textbf{x}$, {$b(\textbf{x},r)$ denotes a disk centered at $\textbf{x}$ with radius $r$}, {and $s(\textbf{x},r,\phi)$ denotes the sector of the disk centered at $\textbf{x}$ with radius $r$ and angle $\phi$; when $\phi=\pi$, we denote the half-disk by $s(\textbf{x},r)$}. $\mathcal{L}_X(s)=\mathbb{E}[e^{-sX}]$ denotes the Laplace transform (LT) of the PDF of the random variable $X$. The ordinary hypergeometric function is denoted by ${}_2F_1$. 

%

\begin{figure}[h]
\begin{minipage}[htb]{0.98\linewidth}
\centering\includegraphics[width=0.75\columnwidth]{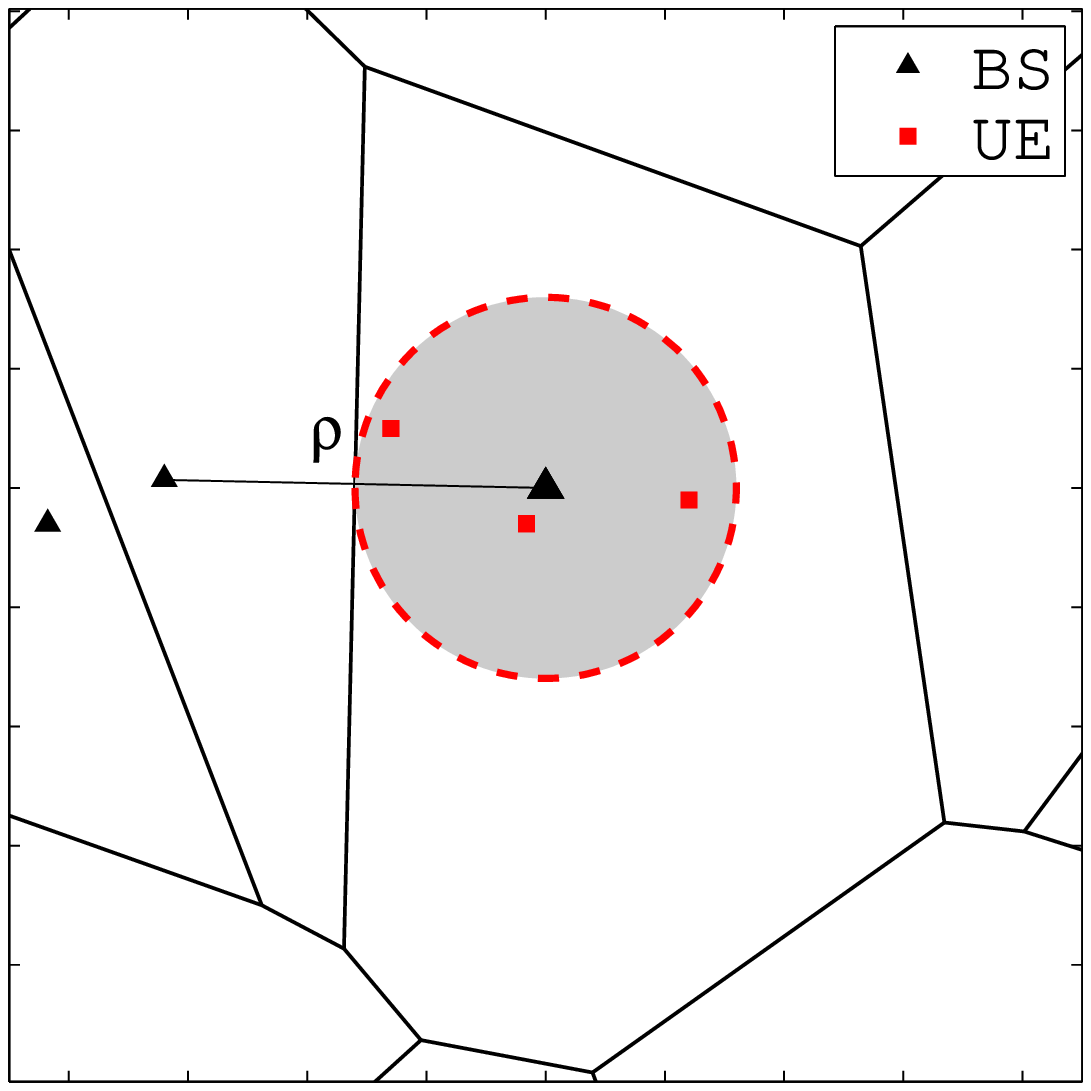}
\subcaption{Model 1: sector {corresponds to} in-disk.}\label{a_sysModelPhi}
\end{minipage}\;\;
\begin{minipage}[htb]{0.98\linewidth}
\centering\includegraphics[width=0.75\columnwidth]{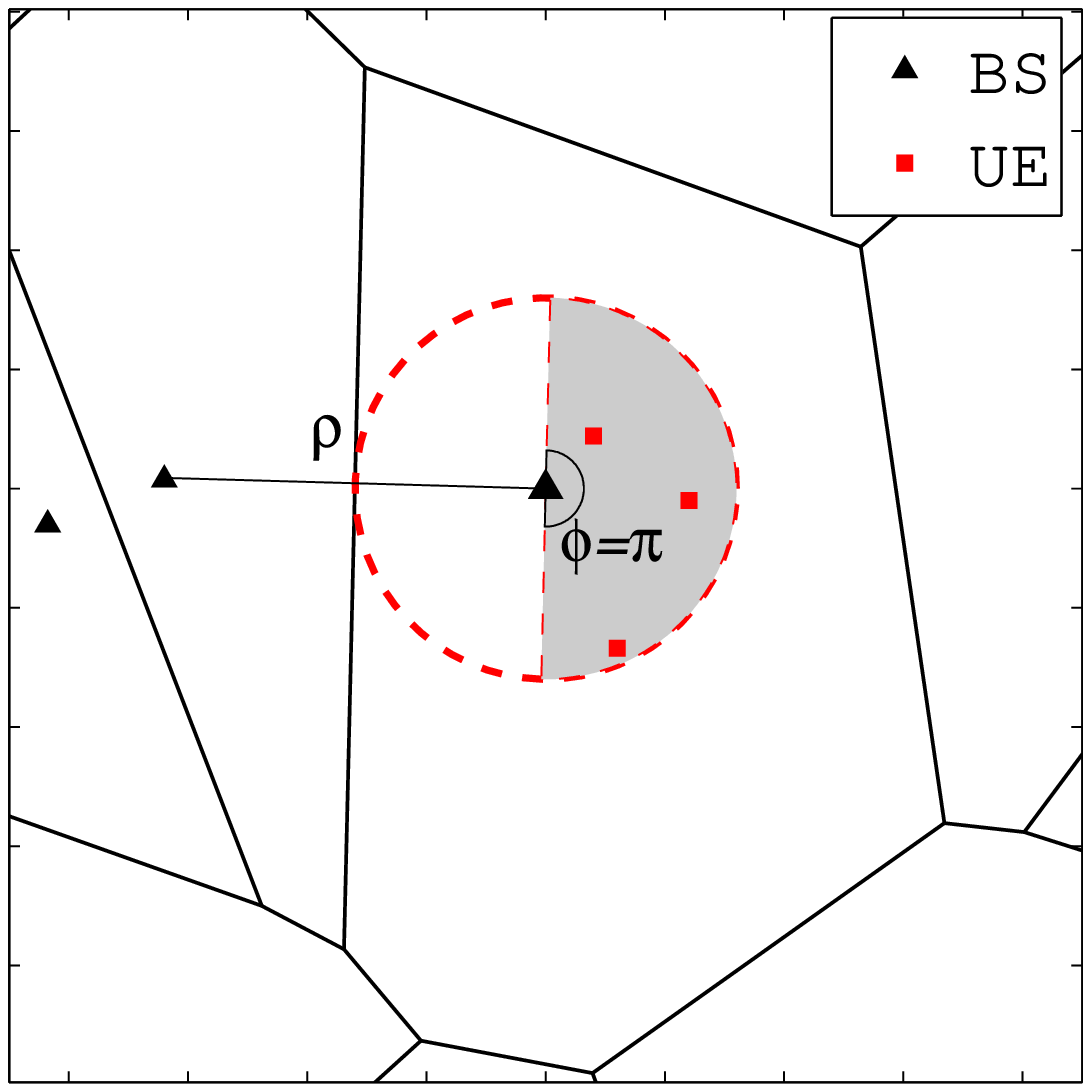}
\subcaption{Model 2: sector {is half of the} in-disk.}\label{b_sysModelPhi}
\end{minipage}
\caption{{A realization of the network with $N=3$ for Models 1 and 2. The UEs, in-disk (dashed circle) and sector (shaded) for the cell at \textbf{o} are shown.}}\label{sysModelPhi}
\end{figure}

\section{System Model}\label{SysMod}
\subsection{NOMA System Model}
We consider a downlink cellular network where BSs transmit with a total power budget of {$P=1$}. Each BS serves $N$ UEs in one time-frequency resource block by multiplexing the signals for each UE with different power levels; here $N$ denotes the cluster size. The BSs use fixed-rate transmissions, {where the rate can be} different for each UE, {referred to as the transmission rate of the UE. Such transmissions lead to effective rates that are lower than the {transmission rate}; we refer to the effective rate of a UE as the throughput of the UE}. The BSs are distributed according to a homogeneous Poisson point process (PPP) $\Phi$ with intensity $\lambda$. To the network we add an extra BS at the origin \textbf{o}, which, under expectation over the PPP $\Phi$, becomes the typical BS serving UEs in the typical cell. In this work we study the performance of the typical cell. Note that since $\Phi$ does not include the BS at \textbf{o}, $\Phi$ is the set of the interfering BSs for the UEs in the typical cell. {Denote by $\rho$ the distance between the BS at \textbf{o} and its nearest neighbor.} Since $\Phi$ is a PPP, the distance $\rho$ follows the distribution
\begin{align}
f_{\rho}(x)=2\pi \lambda x e^{-\pi \lambda x^2}, \;\;\; x\geq 0.\label{f_rho}
\end{align} 
Consider a disk centered at the {\textbf{o}} with radius $\rho/2$. We refer to this as the in-disk as shown in Fig. \ref{sysModelPhi}. The in-disk is the largest disk centered at the serving BS that fits inside the Voronoi cell. UEs outside of this disk are relatively far from their BS, have weaker channels and thus are better served in their own resource block (without sharing) or even using {coordinated multipoint (CoMP) transmission} if they are near the cell edge \cite{comp1,comp2}. These UEs are not discussed further in this work. {We focus on UEs inside the in-disk {since they have good channel conditions, {yet enough disparity among themselves,} and thus can effectively be served using} NOMA.}


{We consider three models {for the clustering of UEs}. Each model {results in} a Poisson cluster process with $N$ points distributed uniformly and independently in each cluster. Let $\textbf{x}$ be the parent point, i.e., the BS, and $\rho_\textbf{x}$ the distance to its nearest neighbor $\textbf{y}_\textbf{x}$; for brevity, $\rho_\textbf{o}$ is denoted by $\rho$. The points in the cluster are: 
\begin{itemize}
\item Model 1: distributed on {the disk} $b(\textbf{x},\rho_\textbf{x}/2)$
\item Model 2: distributed on the half-disk $s(\textbf{x},\rho_\textbf{x}/2)$ such that all points in $s$ have distance at least $\rho_\textbf{x}$ from $\textbf{y}_\textbf{x}$
\item Model 3: distributed on the line segment $s(\textbf{x},\rho_\textbf{x}/2)\cap \ell(\textbf{x},\textbf{y}_\textbf{x})$, where $\ell(\textbf{x},\textbf{y}_\textbf{x})$ is the line through $\textbf{x}$ and $\textbf{y}_\textbf{x}$
\end{itemize}
More compactly, let $s(\textbf{x},\rho_\textbf{x}/2,\phi)\subseteq b(\textbf{x},\rho/2)$ be the {(closed)} disk sector of angle $\phi$ whose curved boundary has midpoint $\textbf{z}_\textbf{x}=(3\textbf{x}-\textbf{y}_\textbf{x})/2$. Then for Model 1, $\phi=2\pi$, for Model 2, $\phi=\pi$, and for Model 3, $\phi=0$.} A realization of the cell at \textbf{o}, its in-disk, and the surrounding cells are shown in Fig. \ref{sysModelPhi}; {the sectors $s(\textbf{x},\rho_\textbf{x}/2,\phi)$ are shown shaded for Models 1 and 2.}


For Model 1, the union of all the disks $\bigcup_{\textbf{x}\in\Phi} b(\textbf{x},\rho_\textbf{x}/2)$ is the so-called {\em Stienen model} \cite{stienen3}. The area fraction covered by the Stienen model is 1/4. This means that if all users form a stationary point process, 1/4 of them are served using NOMA in Model 1 and 1/8 in Model 2 (and 0 in Model 3). More generally, for arbitrary $\phi$, the area fraction is $\phi/(8\pi)$. {Note that the NOMA UEs form a Poisson cluster process where a fixed number of daughter points are placed uniformly at random on disks of random radii. {The radii are correlated since the in-disks of two cells whose BSs are mutual nearest neighbors have the same radius and all disks are disjoint, but given the radii, the $N$ daughter points are placed independently.} Hence, there are three important differences to (advantages over) Matern cluster processes: the number of daughter points is fixed, the disk radius is random, and the disks do not overlap.}


{Focusing on the typical cell, the link distance $R$ between a UE uniformly distributed in the sector of the in-disk $s(\textbf{o},\rho/2,\phi)$ with $\phi>0$ and the BS at \textbf{o}, conditioned on $\rho$, follows
\begin{align}
f_{R\mid \rho}(r \mid \rho)=\frac{8r}{\rho^2}, \;\;\; 0 \leq r \leq \frac{\rho}{2}. \label{f_R}
\end{align}
Since $\phi>0$ for Models 1 and 2, the statistics of their link distances are according to \eqref{f_R}. For Model 3, however, the sector becomes a line segment as $\phi \rightarrow 0$. Consequently, $R$, conditioned on $\rho$, in Model 3 follows the distribution
\begin{align}
f_{R\mid \rho}(r \mid \rho)=\frac{2}{\rho}, \;\;\; 0 \leq r \leq \frac{\rho}{2}. \label{f_R_Mod4}
\end{align} 
\emph{\textbf{Remark 1:}} {Given $\rho$, the exact distance between the UE and the interferer nearest to \textbf{o} in Model 3 is $z=R + \rho$.}\\
\emph{\textbf{Remark 2:}} As there is no interfering BS inside $b(\textbf{o},\rho)$, a UE located at $\textbf{u}$ in $s(\textbf{o},\rho/2,\phi)$, for any $\phi$, is $\rho-R$ away from the boundary of this disk. Hence, all three models guarantee that there is no interfering BS in $b(\textbf{u},\rho-R)$.}


It makes sense to employ NOMA for UEs that have good channel conditions and thus can afford to share resource blocks with other UEs rather than those that cannot. Accordingly, any user close to a cell edge is worse off than the cell center users, on average. As $\phi$ decreases, users are located in the in-disk farther from any cell edge, particularly the edge closest to \textbf{o}, and consequently have better intercell interference conditions. {In this context, Model 2 can be used as a technique to improve the performance by selecting UEs for NOMA operation, i.e., UE clustering, more efficiently based on their locations, and Model 3 can be viewed as a method to upper bound the achievable performance via UE clustering.} 




A Rayleigh fading environment is assumed such that the fading coefficients {are i.i.d.~with} a unit mean exponential distribution. A power law path loss model is considered where the signal power decays at the rate $r^{-\eta}$ with distance $r$, where $\eta>2$ denotes the path loss exponent and $\delta=\frac{2}{\eta}$. 

SIC is employed for decoding NOMA UEs. According to the NOMA scheme, the PA and transmission rate are designed such that the $i^{\rm th}$ strongest UE is able to decode the messages intended for all those UEs weaker than itself. This requires ordering of UEs based on the quality of the transmission link. We order UEs in such a way that the $i^{\rm th}$ UE, referred to as $\text{UE}_i$, has the $i^{\rm th}$ strongest transmission link. There are various ways to define what comprises the link quality. The link quality should include the effect of path loss (and therefore link distance), fading and intercell interference. The impact of the large-scale path loss is more stable and dominant than the fading effect which varies on a much shorter time scale. Additionally, accounting for intercell interference and fading necessitates very high feedback overhead. Since for small values of $N$ the path loss dominates the channel relative to fading, considering the quality of a channel to be based on the distance between a UE and its BS is often assumed to be a reasonable approximation \cite{N2,N13,N4,N6,N6_20,N6_20_59}. The link quality can be determined by ordering the UEs of the typical cell {from strongest to weakest} according to descending
\begin{itemize}
\item Mean signal power (MSP)\footnote{{It should be noted that this ordering is based on the total unit power transmission received at the UE.}}: this ignores fading and therefore orders UEs based on descending $R^{-\eta}$. Equivalently it can be viewed as ordering based on ascending link distance $R$.
\item Instantaneous signal power (ISP): this includes fading and therefore orders UEs based on descending $hR^{-\eta}$.
\item {Mean-fading signal-to-intercell-interference-and-noise ratio (MFS$\oset[-.1ex]{_{\o}}{\raise.01ex\hbox{I}}$NR)}: this {assumes channels with the mean fading gain of 1} in both the transmission from the serving BS and in the intercell interference and therefore orders UEs based on descending $\frac{R^{-\eta}}{\sum_{\mathbf{x \in \Phi}} \|\mathbf{x}-\mathbf{u}\|^{-\eta} + \sigma^2}$ where $ \|\mathbf{x}\|$ and $\|\mathbf{u}\|$ are the locations of the interfering BSs and UE, respectively and $\sigma^2$ is the noise power.
\item Instantaneous signal-to-intercell-interference-and-noise ratio (IS$\oset[-.1ex]{_{\o}}{\raise.01ex\hbox{I}}$NR): this includes fading and therefore orders UEs based on descending $\frac{h R^{-\eta}}{I^{\rm\o} + \sigma^2}$.
\end{itemize}
{Analyzing the SINR for ordering based on ISP and {MFS$\oset[-.1ex]{_{\o}}{\raise.01ex\hbox{I}}$NR} is out of the scope of this work.} Hence, we analyze and compare the following two schemes:
\begin{itemize}
\item MSP-based: the UEs of the typical cell are indexed according to their ascending ordered distance $R_i$; the $i^{\rm th}$ closest UE from \textbf{o} is referred to as $\text{UE}_i$, for $1 \leq i \leq N$.
\item {IS$\oset[-.1ex]{_{\o}}{\raise.01ex\hbox{I}}$NR-based: UEs of the typical cell are indexed with respect to their descending ordered IS$\oset[-.1ex]{_{\o}}{\raise.01ex\hbox{I}}$NR $Z_i$\footnote{Note that $Z_i$ is equivalent to ${\rm SINR}_{\rm OMA}^i$ in \eqref{realSINR_OMA}. We use the notation $Z_i$ for brevity and to differentiate between the context it is being used in.}; hence, $\text{UE}_i$ has the $i^{\rm th}$ largest IS$\oset[-.1ex]{_{\o}}{\raise.01ex\hbox{I}}$NR, for $1 \leq i \leq N$.}
\end{itemize}   The power for the signal intended for $\text{UE}_i$ is denoted by $P_i$, hence $P=\sum_{i} P_i$.


To successfully decode its own message, $\text{UE}_i$ must therefore be able to decode the messages intended for all UEs weaker than itself, i.e, $\text{UE}_{i+1},\dots,\text{UE}_N$. This is achieved by allocating higher powers and/or lower transmission rates to the data streams of the weaker UEs. Correspondingly, $\text{UE}_i$ is not able to decode any of the streams sent to UEs stronger than itself, i.e., $\text{UE}_1,\dots,\text{UE}_{i-1}$ due to their smaller powers and/or higher transmission rates. Assuming perfect SIC, {the intraference experienced at $\text{UE}_i$ {when decoding its own message}, $I^\circ_i$, is comprised of} the powers from the messages intended for $\text{UE}_1,\dots,\text{UE}_{i-1}$. Since in practice SIC is not perfect, our mathematical model additionally considers a fraction $0 \leq \beta \leq 1$ of RI from the UEs weaker than $\text{UE}_i$ in $I^\circ_i$ {in a fashion similar to \cite{noma_sic}}. When perfect SIC is assumed, $\beta=0$, while $\beta=1$ corresponds to no SIC at all. Additionally, $\text{UE}_i$ suffers from intercell interference, $I^{\rm\o}_i$, arising from the power received from all the other BSs in the network, and noise power $\sigma^2$. For the NOMA network $2N-1$ parameters are to be selected, namely $N$ transmission rates and $N-1$ powers. {Note that MSP-based ordering of UEs is agnostic to intercell interference and fading; however, our RA (choice of the $2N-1$ parameters) is not. For the case of IS$\oset[-.1ex]{_{\o}}{\raise.01ex\hbox{I}}$NR-based ordering, both ordering and RA are intercell interference- and fading-aware.}


\subsection{OMA System Model}
We compare our NOMA model with a traditional OMA network where only one UE is served by each BS in a single time-frequency resource block. We focus on time division multiple access (TDMA). For a fair comparison with the NOMA system, the BS serves $N$ UEs distributed uniformly at random in {(part of)} the in-disk as in the NOMA setup {according to the model being employed}. Each TDMA message is transmitted using full power $P=1$ for a duration $T_i$. Without loss of generality, a unit time duration is assumed for a NOMA transmission and therefore $\sum_i T_i=1$. Consequently, $2N-1$ parameters are to be selected for the OMA network, namely $N$ transmission rates and $N-1$ fractions of the time slot. We compare both MSP-based UE ordering and IS$\oset[-.1ex]{_{\o}}{\raise.01ex\hbox{I}}$NR-based ordering for the OMA model, too. 

\section{SINR Analysis}

\subsection{SINR in NOMA Network}

In the NOMA network, the SINR at $\text{UE}_i$ of {the message intended for $\text{UE}_j$} in the typical cell for $i \leq j \leq N$ is 
\begin{align*}
{\rm SINR}_j^i=\!\!\frac{h_i R_i^{-\eta} P_j}{ \underbrace{h_i R_i^{-\eta}  \Bigg( \sum\limits_{m=1}^{j-1} P_m \!\!+ \beta \!\! \sum\limits_{k=j+1}^{N} \!\! P_k \Bigg)}_{{I^\circ_{j,i}}} \!\! +  \!\! \underbrace{\sum\limits_{\textbf{x} \in {\Phi}}  g_{\textbf{y}_i} {\|\textbf{y}_i\|}^{-\eta}}_{I^{\rm\o}_i} + \sigma^2},
\end{align*}
where $\textbf{y}_i=\textbf{x}-\textbf{u}_i$, $\textbf{u}_i$ is the location of $\text{UE}_i$, $h_i$ ($g_{\textbf{y}_i}$) is the fading coefficient from the serving BS (interfering BS) located at $\textbf{o}$ ($\textbf{x}$) to $\text{UE}_i$. The intraference experienced when $\text{UE}_i$ decodes the message for $\text{UE}_j$ is denoted by $I^\circ_{j,i}$. We use $I^\circ_{i}$ to denote $I^\circ_{i,i}$.

\subsection{Laplace Transform of the Intercell Interference}
We analyze the LT of the intercell interference at both the unordered UE and the UE ordered based on MSP. {Upon taking the expectation over the BS PPP and the {unordered} UE's {(ordered $\text{UE}$'s)} location, the UEs in the cell with the BS at $\textbf{o}$ become the typical {unordered} UEs {(typical ordered UEs, from $\text{UE}_1$ to $\text{UE}_N$.)}.}

\textbf{\emph{Lemma 1:}} The LT of $I^{\rm\o}$ {($I^{\rm\o}_i$)} at the typical unordered $\text{UE}$ {(ordered $\text{UE}_i$)} conditioned on $R$  {($R_i$)} and $\rho$, where $u=\rho-R$  {($u_i=\rho-R_i$)}, {in Model 1} is approximated as
\begin{align}
\mathcal{L}_{I^{\rm\o} \mid R,\rho} (s)  &\approx  \exp \left({   - \frac{2 \pi \lambda s }{(\eta-2){u}^{\eta-2}} { }_2F_1  \left( 1,1 \! - \! \delta; 2\! - \! \delta; \frac{-s}{{u}^{\eta}} \right)   } \right) \nonumber \\
& \;\;\; \times  \frac{1}{1+s \rho^{-\eta}}  \label{L_I_unordered}\\
&\stackrel{\eta=4}= e^{-\pi \lambda \sqrt{s} \tan^{-1} \left(\frac{\sqrt{s}}{u^{2}} \right)}  \frac{1}{1+s \rho^{-4}}\label{L_I4_unordered}
\end{align} 
\begin{align}
{\mathcal{L}_{I^{\rm\o}_i \mid R_i,\rho}} (s)  &\approx  \exp \left({   - \frac{2 \pi \lambda s }{(\eta-2){{u_i}}^{\eta-2}} { }_2F_1  \left( 1,1 \! - \! \delta; 2\! - \! \delta; \frac{-s}{u_i^{\eta}} \right)   } \right) \nonumber \\
& \;\;\; \times  \frac{1}{1+s \rho^{-\eta}}  \label{L_I}\\
&\stackrel{\eta=4}= e^{-\pi \lambda \sqrt{s} \tan^{-1} \left(\frac{\sqrt{s}}{{u_i}^{2}} \right)}  \frac{1}{1+s \rho^{-4}}.\label{L_I4}
\end{align}
\begin{IEEEproof}
Let $\textbf{y}=\textbf{x}-\textbf{u}$, where $ \|\mathbf{x}\|$ and $\|\mathbf{u}\|$ are the locations of the interfering BSs and the UE, respectively. The fading coefficient from the interfering BS at $\textbf{x}$ to the UE is $g_{\textbf{y}}$. The intercell interference experienced at the UE is 
\begin{align}
I^{\rm\o} =  \sum_{\substack{\textbf{x}\in\Phi\\ \|\textbf{x}\|>\rho}} g_{\textbf{y}} {\|\textbf{y}\|}^{-\eta} +  \sum_{\substack{\textbf{x}\in\Phi\\ \|\textbf{x}\|=\rho}}  g_{\textbf{y}} \|\textbf{y}\|^{-\eta} . \label{I_inter_rewrite_unordered}
\end{align}
The first term of the LT accounts for the first term in \eqref{I_inter_rewrite_unordered} corresponding to the non-nearest interferers from $\textbf{o}$ lying at a distance at least $u$ {($u_i$)} from the unordered UE {(ordered $\text{UE}_i$)}. It is obtained from employing Slivnyak's theorem, the probability generating functional of the PPP, and MGF of $g_{\textbf{y}}  \sim \exp(1)$. However, this does not include the BS at distance $\rho$ from \textbf{o}, which is accounted for by the second term in \eqref{I_inter_rewrite_unordered} using the MGF of $g_{\textbf{y}}$. Denote by $z$ the distance between this interferer and the typical UE. {Then using the MGF of $g_{\textbf{y}} $, the exact expression for the LT of the second term in \eqref{I_inter_rewrite_unordered}} is $\mathbb{E}_{z \mid \rho} \left[ \left(1+s z^{-\eta}\right)^{-1} \right]$. For simplicity we approximate it using the approximate mean of this distance. Since the average position of the typical UE {distributed uniformly in the in-disk} is \textbf{o}, $\mathbb{E}[z \mid \rho] \approx \rho$. 
\end{IEEEproof}

\emph{Note:} The first term of the LT of $I^{\rm\o}$ ($I^{\rm\o}_i$) is pessimistic since the interference guard zone in our model $u$ ($u_i$) is smaller than the actual one. For the second term, an exact evaluation {(by simulation)} shows that the difference between $\mathbb{E}[z \mid \rho]$ and $\rho$ is less than 3.2\%.

{In the case of Model 2 the distance between the UEs and the interferer closest to $\textbf{o}$ is larger than in the case of Model 1. This corresponds to a change in the impact of the second term of \eqref{I_inter_rewrite_unordered}. The LT of intercell interference changes accordingly.\\
\textbf{\emph{Lemma 2:}} {The LT of $I^{\rm\o}$ ($I^{\rm\o}_i$) at the typical unordered $\text{UE}$ (ordered $\text{UE}_i$) conditioned on $R$ ($R_i$) and $\rho$, where $u=\rho-R$ ($u_i=\rho-R_i$), in Model 2 is approximated as
\begin{align}
\mathcal{L}_{I^{\rm\o} \mid R,\rho} (s)  &\approx  \exp \left({   - \frac{2 \pi \lambda s }{(\eta-2){u}^{\eta-2}} { }_2F_1  \left( 1,1 \! - \! \delta; 2\! - \! \delta; \frac{-s}{{u}^{\eta}} \right)   } \right) \nonumber \\
& \;\;\; \times \frac{1}{1+s (1.25\rho)^{-\eta}}  \label{L_I_unordered_Mod3}\\
&\stackrel{\eta=4}= e^{-\pi \lambda \sqrt{s} \tan^{-1} \left(\frac{\sqrt{s}}{u^{2}} \right)}  \frac{1}{1+s (1.25\rho)^{-4}}\label{L_I4_unordered_Mod3}
\end{align} 
\begin{align}
\mathcal{L}_{I^{\rm\o}_i \mid R,\rho} (s)  &\approx  \exp \left({   - \frac{2 \pi \lambda s }{(\eta-2){u_i}^{\eta-2}} { }_2F_1  \left( 1,1 \! - \! \delta; 2\! - \! \delta; \frac{-s}{{u_i}^{\eta}} \right)   } \right) \nonumber \\
& \;\;\; \times \frac{1}{1+s (1.25\rho)^{-\eta}} \label{L_I_ordered_Mod3}\\
&\stackrel{\eta=4}= e^{-\pi \lambda \sqrt{s} \tan^{-1} \left(\frac{\sqrt{s}}{u_i^{2}} \right)}  \frac{1}{1+s (1.25\rho)^{-4}}.\label{L_I4_ordered_Mod3}
\end{align} }
\begin{IEEEproof}
The proof follows according to that in Lemma 1. However, in the second term, $\mathbb{E}[z \mid \rho] \approx 1.25\rho$. {We use this approximation because a UE located in Model 2, i.e. in the half-disk away from the interferer nearest to \textbf{o}, has $\rho \leq \mathbb{E}[z \mid \rho] \leq 1.5\rho$; consequently, we approximate the average position of a UE in this model and $z$ accordingly. An exact evaluation {(by simulation)} for Model 2 shows that the difference between $\mathbb{E}[z \mid \rho]$ and $1.25\rho$ is less than 0.92\%.}
\end{IEEEproof}}

{In the case of Model 3 the distance between the UEs and the interferer closest to $\textbf{o}$ is exactly $z=R+\rho$. This too corresponds to a change in the impact of the second term of \eqref{I_inter_rewrite_unordered}. The LT of intercell interference changes accordingly.\\
\textbf{\emph{Lemma 3:}} The LT of $I^{\rm\o}$ ($I^{\rm\o}_i$) at the typical unordered $\text{UE}$ (ordered $\text{UE}_i$) distributed according to Model 3, conditioned on $R$ ($R_i$) and $\rho$, where $u=\rho-R$ ($u_i=\rho-R_i$), $a_1=\frac{(1.5\rho)^{\eta}}{s}$ and $a_2=\frac{\rho^{\eta}}{s}$, is {approximated as}
\begin{align}
&\mathcal{L}_{I^{\rm\o} \mid R,\rho} (s)  {\approx}  \exp \left({ \!  - \frac{2 \pi \lambda s }{(\eta-2){u}^{\eta-2}} { }_2F_1 \! \left( \! 1,1 \! - \! \delta; 2\! - \! \delta; \frac{-s}{{u}^{\eta}} \! \right) \!  } \right) \times   \nonumber\\
& \left( \! 1 \!-\! 3  \;  { }_2F_1 \! \left( \! 1,\frac{1}{\eta} ; \frac{\eta+1}{\eta}; -a_1 \! \right)   \! + \! 2 \;  { }_2F_1  \! \left( \! 1,\frac{1}{\eta} ; \frac{\eta + 1}{\eta}; -a_2 \! \right) \! \right)  \label{L_I_unordered_Mod4}   \\
&\stackrel{\eta=4}= e^{-\pi \lambda \sqrt{s} \tan^{-1} \left(\frac{\sqrt{s}}{u^{2}} \right)} \times \Bigg(  \! 1 \!-\!   \frac{\tan^{-1} \! \left(a_1^{\frac{1}{4}}   \right)  +  \tanh^{-1} \! \left( a_1^{\frac{1}{4}}   \right)}{ \frac{2}{3} a_1^{\frac{1}{4}}  }   \nonumber \\
&  \;\;\;\; +  \frac{\tan^{-1} \! \left( a_2^{\frac{1}{4}}  \right)  +  \tanh^{-1} \! \left( a_2^{\frac{1}{4}}   \right)}{   a_2^{\frac{1}{4}}   } \! \Bigg)\label{L_I4_unordered_Mod4}
\end{align}
\begin{align}
&\mathcal{L}_{I^{\rm\o}_i \mid R,\rho} (s)  {\approx}  \exp \left({ \!  - \frac{2 \pi \lambda s }{(\eta-2){u_i}^{\eta-2}} { }_2F_1 \! \left( \! 1,1 \! - \! \delta; 2\! - \! \delta; \frac{-s}{{u_i}^{\eta}} \! \right) \!  } \right) \times \nonumber \\
& \left( \! 1 \!-\! 3  \;  { }_2F_1 \! \left( \! 1,\frac{1}{\eta} ; \frac{\eta+1}{\eta}; -a_1 \! \right)   \! +    2 \;  { }_2F_1  \! \left( \! 1,\frac{1}{\eta} ; \frac{\eta + 1}{\eta}; -a_2 \! \right) \! \right)  \label{L_I_ordered_Mod4}   \\
&\stackrel{\eta=4}= e^{-\pi \lambda \sqrt{s} \tan^{-1} \left(\frac{\sqrt{s}}{u_i^{2}} \right)} \times \Bigg(  \! 1 \!-\!   \frac{\tan^{-1} \! \left( a_1^{\frac{1}{4}}   \right)  +  \tanh^{-1} \! \left( a_1^{\frac{1}{4}}   \right)}{ \frac{2}{3} a_1^{\frac{1}{4}}  }    \nonumber \\
& \;\;\;\; +  \frac{\tan^{-1} \! \left( a_2^{\frac{1}{4}} \right)  +  \tanh^{-1} \! \left( a_2^{\frac{1}{4}}   \right)}{   a_2^{\frac{1}{4}}   } \! \Bigg).\label{L_I4_ordered_Mod4}
\end{align}
\begin{IEEEproof}
The first term of \eqref{L_I_unordered_Mod4} and \eqref{L_I_ordered_Mod4} follows from the first term of the LTs in Lemma 1. The exact second term is $\mathbb{E}_{z \mid \rho} \left[ \left(1+s z^{-\eta}\right)^{-1} \right]$. Since $z=R+\rho$, using \eqref{f_R_Mod4}, $f_{z\mid\rho}(y \mid \rho)=2/\rho, \; \rho \leq y \leq 3\rho/2$,
\begin{align*}
&\mathbb{E}_{z \mid \rho} \left[ \left(1+s z^{-\eta}\right)^{-1} \right]= \int_{\rho}^{1.5 \rho} \frac{1}{1+s y^{-\eta}} f_{z\mid\rho}(y \mid \rho) dy  \\
&= 1 \!-\! 3  \;  { }_2F_1 \! \left( \! 1,\frac{1}{\eta} ; \frac{\eta+1}{\eta}; -a_1 \! \right)   \! + \!  2 \;  { }_2F_1  \! \left( \! 1,\frac{1}{\eta} ; \frac{\eta + 1}{\eta}; -a_2 \! \right) \!  .
\end{align*}
\end{IEEEproof}

\subsection{UE Ordering}
Since the order of the UEs is known at the BS, we use order statistics for the PDFs of the link quality. {These are derived using the distribution of the unordered link quality statistics and the theory of order statistics \cite{N16_18}.}

\subsubsection{MSP-Based Ordering}
{UEs are ordered  based on the ascending ordered link distance $R_i$. Hence, $R_i$ is the distance between the $i^{\rm th}$ nearest UE, i.e., $\text{UE}_i$ to its serving BS, given $\rho$, for $1 \leq i \leq N$. Using the distribution of the unordered link distance $R$ conditioned on $\rho$ in \eqref{f_R} {for Models 1 and 2} we have
\begin{align}
f_{R_i \mid \rho}(r \mid  \rho) =\binom{N-1}{i-1}\frac{8rN }{\rho^2} \!  \left( \frac{4r^2}{\rho^2} \right)^{i-1} \!\! \left( 1 \!-\!\frac{4r^2}{\rho^2}\right)^{N-i}\!\! \label{f_Ri}
\end{align}
for $0 \leq r \leq  {\rho}/{2}$, where $\binom{c}{d}=\frac{c!}{d! (c-d)!}$.

{{Similarly,} using the distribution of the unordered link distance $R$ conditioned on $\rho$ in \eqref{f_R_Mod4} for Model 3 we have
\begin{align}
f_{R_i \mid \rho}(r \mid  \rho) =\binom{N-1}{i-1}N\frac{2 }{\rho}  \left( \frac{2r}{\rho} \right)^{i-1} \left( 1 \!-\!\frac{2r}{\rho}\right)^{N-i}\!\! \label{f_Ri_Mod4}
\end{align}
for $0  \leq r \leq  {\rho}/{2}$.
}

{Note that MSP-based ordering guarantees that the nearest interfering BS from $\text{UE}_i$ is farther than $\rho-R_i$.}}

\subsubsection{IS$\oset[-.1ex]{_{\o}}{\raise.01ex\hbox{I}}$NR-Based Ordering}
{UEs are ordered based on descending ordered IS$\oset[-.1ex]{_{\o}}{\raise.01ex\hbox{I}}$NR, $Z_i$. The unordered IS$\oset[-.1ex]{_{\o}}{\raise.01ex\hbox{I}}$NR is $Z=\frac{hR^{-\eta}}{{I^{\rm\o}}+\sigma^2}$.}

\textit{\textbf{Lemma 4:}} The CDF of the unordered IS$\oset[-.1ex]{_{\o}}{\raise.01ex\hbox{I}}$NR $Z$ conditioned on $\rho$ is approximated as
\begin{align}
F_{Z \mid \rho}(x)\!&\approx\!1- \!\! \int_{0}^{\rho/2} \!\! \!\!\mathcal{L}_{I^{\rm\o} \mid R, \rho} (x r^{\eta}) \exp(-x r^{\eta} \sigma^2 ) f_{R \mid \rho}(r) dr, \label{F_Z}
\end{align}
where $\mathcal{L}_{I^{\rm\o} \mid R, \rho}(s)$ {is approximated in Lemmas 1, 2, and 3 for Models 1, 2, and 3, respectively, and $f_{R \mid \rho}(r)$ is given in \eqref{f_R} for Models 1 and 2 and in \eqref{f_R_Mod4} for Model 3.}
\begin{IEEEproof}
By definition of $Z$,
\begin{align*}
F_{Z \mid \rho}(x)&=\mathbb{E}_{R,{I^{\rm\o}}} \left[ \mathbb{P} \left( h \leq x R^{\eta} (I^{\rm\o} + \sigma^2) \mid R,{I^{\rm\o}} \right) \right] \nonumber \\
&\stackrel{(a)}= \mathbb{E}_{R,{I^{\rm\o}}} \left[ 1 - \exp(-x R^{\eta} I^{\rm\o}) \exp(-x R^{\eta} \sigma^2 )   \right] \nonumber \\
&\stackrel{(b)}\approx 1- \int_{0}^{\rho/2} \mathcal{L}_{I^{\rm\o} \mid R, \rho} (x r^{\eta}) \exp({-x r^{\eta} \sigma^2}) f_{R \mid \rho}(r) dr.
\end{align*}
Here (a) follows from $h\sim\exp(1)$ and (b) using the definition of the LT of $I^{\rm\o}$ conditioned on $R$ and $\rho$. 
\end{IEEEproof}
\textit{\textbf{Corollary 1:}} The CDF of the ordered IS$\oset[-.1ex]{_{\o}}{\raise.01ex\hbox{I}}$NR $Z_i$ conditioned on $\rho$ is approximated as
\begin{align}
F_{Z_i \mid \rho}(x) \approx \!\!\!\! \sum_{k=N+1-i}^N \!\! \binom{N}{k} \!\! \left( F_{Z \mid \rho}(x) \right)^{k} \! \left(1- F_{Z \mid \rho}(x) \right)^{N-k}\!\!,
\end{align}
where $F_{Z \mid \rho}(x)$ is given in \eqref{F_Z}.

\subsection{Coverage in NOMA Network}
In order to decode its intended message, $\text{UE}_i$ needs to decode the messages intended for all UEs weaker than itself. We use $\theta_j$ to denote the SINR threshold corresponding to the transmission rate associated with the message for $\text{UE}_j$. {Coverage at $\text{UE}_i$ is defined as the event 
\begin{align}
&C_i=  \bigcap\limits_{j=i}^N \left\lbrace {\rm SINR}_j^i\!>\! \theta_j \right\rbrace = \bigcap\limits_{j=i}^N \left\lbrace  h_i  >  R_i^{\eta} (I^{\rm\o}_i \!+\! \sigma^2) \frac{\theta_j}{\tilde{P}_j}\right\rbrace , \label{C_i_new}
\end{align}
where {$\tilde{P}_j= P_j - \theta_j \left( \sum\limits_{m=1}^{j-1} P_m + \beta \!\! \sum\limits_{k=j+1}^{N} \!\! P_k \right).$}\\
\textbf{\textit{Remark 3:}} We observe that the impact of the intraference is that of a reduction in the effective transmit power to $\tilde{P}_j$; without intraference, $\tilde{P}_j$ in \eqref{C_i_new} would be replaced by $P_j$. This reduction and thus $\tilde{P}_j$ is dependent on the transmission rate of the message to be decoded.

We introduce the notion of \emph{NOMA necessary condition} for coverage, which is coverage when only intraference, arising from NOMA UEs within a cell, is considered. By definition we can write the signal-to-intraference ratio (S$\oset[-.1ex]{_{\circ}}{\raise.01ex\hbox{I}}$R) of the message for $\text{UE}_j$ at $\text{UE}_i$ as
\begin{align}
 \text{S}\overset{_\circ}{\rm I}\text{R}_j^i&=\!\! \frac{h_i R_i^{-\eta} P_j}{ {   \frac{h_i}{ R_i^{\eta}} \Bigg(\! \sum\limits_{m=1}^{j-1} P_m \!\! + \!\!\beta \!\! \sum\limits_{k=j+1}^{N} \!\! P_k \! \Bigg)} \!\! } =\frac{ P_j }{   \sum\limits_{m=1}^{j-1} P_m \!\! + \!\! \beta \!\! \sum\limits_{k=j+1}^{N} \!\! P_k  }. \label{genNOMAcover} 
\end{align}
From \eqref{genNOMAcover}, the S$\oset[-.1ex]{_{\circ}}{\raise.01ex\hbox{I}}$R of the message for $\text{UE}_j$ is independent of the UE (i.e., $\text{UE}_i$) it is being decoded at; hence, it can be rewritten as $\text{S$\oset[-.1ex]{_{\circ}}{\raise.01ex\hbox{I}}$R}_j$. In order for the message of $\text{UE}_j$ to satisfy the NOMA necessary condition for coverage, we require
\begin{align}
  \text{S$\oset[-.1ex]{_{\circ}}{\raise.01ex\hbox{I}}$R}_j > \theta_j  \quad \Leftrightarrow \quad \tilde{P}_j>0. \label{indivThet}
\end{align}
The above condition constrains the transmission rate for the message of $\text{UE}_j$ to be less than a certain value that is a function of the power distribution among the NOMA UEs. If this condition is not satisfied, the message for $\text{UE}_j$ cannot be decoded since $\text{S$\oset[-.1ex]{_{\circ}}{\raise.01ex\hbox{I}}$R}_j$ is an upper bound on ${\rm SINR}_j^i$, $j\geq i$. Consequently $\text{UE}_i$ will be in outage as $\tilde{P}_j$ will not be positive. Note that for the particular case of $\text{UE}_1$ with perfect SIC (i.e., $\beta=0$), there is no intraference and $\text{S$\oset[-.1ex]{_{\circ}}{\raise.01ex\hbox{I}}$R}_1=\infty$ implying $\text{UE}_1$ always satisfies the NOMA necessary condition for coverage when SIC is perfect; equivalently, when $\beta=0$, $\tilde{P}_1=P_1$. Hence, failing to satisfy the NOMA necessary condition for coverage guarantees outage for all UEs that need to decode that message simply because the transmission rate is too high for the given PA. This shows the importance of RA in terms of PA and transmission rate choice.


Using $M_i=\max\limits_{i \leq j \leq N} \frac{\theta_j }{\tilde{P}_j }$, $C_i$ in \eqref{C_i_new} can be rewritten compactly as
\begin{align}
C_i= \{ h_i> R_i^{\eta} (I^{\rm\o}_i + \sigma^2) M_i \}.
\end{align}
Next, we derive the coverage probabilities for UEs using each ordering technique.
\subsubsection{Coverage for UEs Ordered Based on MSP}$\;$

\textbf{\emph{Theorem 1:}} If $\tilde{P}_j>0$, the coverage probability of {the typical $\text{UE}_i$ ordered based on MSP, is approximated as} 
\begin{align}
&\mathbb{P} (C_i)\! \approx \!\!\! \int\limits_0^{\infty}  \int\limits_0^{x/2} \!\! e^{-r^{\eta} \sigma^2  M_i} \mathcal{L}_{I^{\rm\o}_i \mid R_i,\rho } \left({r^{\eta}  M_i} \right)  f_{R_i\mid \rho}(r \mid x )  dr f_{\rho}(x) dx, \label{sinr_2}
\end{align}
where $f_{\rho}(x)$ is given in \eqref{f_rho}, {$f_{R_i\mid \rho}(r \mid x)$ in \eqref{f_Ri} for Models 1 and 2 and in \eqref{f_Ri_Mod4} for Model 3, and $\mathcal{L}_{I^{\rm\o}_i \mid R_i,\rho }$ is approximated in Lemmas 1, 2, and 3 for Models 1, 2, and 3, respectively.} 

\begin{IEEEproof}
\begin{align*}
\mathbb{P}(C_i)&= \mathbb{E}_{\rho} \left[\mathbb{E}_{R_i}  \left[ e^{- R_i^{\eta} \sigma^2 M_i}  \mathbb{E} \left[ e^{-\left( R_i^{\eta} M_i \right) I^{\rm\o}_i }  \mid R_i, \rho \right]   \right] \right],
\end{align*}
as $h_i \sim \exp(1)$. The inner expectation is the conditional LT of $I^{\rm\o}_i$ (given $R_i$ and $\rho$). From this we obtain \eqref{sinr_2}.
\end{IEEEproof}

\subsubsection{Coverage for UEs Ordered Based on IS$\oset[-.1ex]{_{\o}}{\raise.01ex\hbox{I}}$NR}
$\;$ 

\textbf{\textit{Theorem 2:}} If $\tilde{P}_j>0$, the coverage probability of {the typical $\text{UE}_i$ ordered based on IS$\oset[-.1ex]{_{\o}}{\raise.01ex\hbox{I}}$NR, is approximated as} 
\begin{align}
\mathbb{P} (C_i) &\approx   \int_0^{\infty} \left(1- F_{Z_i \mid \rho}(M_i \mid x) \right) f_{\rho}(x) dx, \label{covg_CB}
\end{align}
where $f_{\rho}(x)$ is given in \eqref{f_rho}.
\begin{IEEEproof}
\eqref{covg_CB} follows using $\mathbb{P} (C_i)=\mathbb{P} \left( Z_i > M_i  \right)$.
\end{IEEEproof}

For a given SINR threshold $\theta_i$, corresponding to a transmission (normalized) rate of $\log(1+\theta_i)$, the throughput of the typical $\text{UE}_i$ is
\begin{align}
\mathcal{R}_i=\mathbb{P} (C_i) \log(1+\theta_i). \label{R_i_NOMA}
\end{align}
The cell sum rate is $\mathcal{R}_{\rm tot}=\sum\limits_{i=1}^N \mathcal{R}_i $. 

\subsection{OMA Network}
The SINR for $\text{UE}_i$ of the typical cell is 
\begin{align}
{\rm SINR}_{\rm OMA}^i=\frac{h_i  R_i^{-\eta}}{   \underbrace{\sum\limits_{\textbf{x} \in \Phi}  g_{\textbf{y}_i} {\|\textbf{x}-\textbf{u}_i\|}^{-\eta}}_{I^{\rm\o}_i} + \sigma^2}.\label{realSINR_OMA}
\end{align}
where $\textbf{u}_i$ is the location of $\text{UE}_i$, $h_i$ ($g_{\textbf{y}_i}$) is the fading coefficient from the serving BS (interfering BS) located at $\textbf{o}$ ($\textbf{x}$) to $\text{UE}_i$. {Coverage at $\text{UE}_i$ is defined as $\tilde{C}_i=\left\lbrace {\rm SINR}_{\rm OMA}^i > \theta_i \right\rbrace$.}

\textbf{\emph{Lemma 5:}} In the OMA network, the coverage probability of {the typical $\text{UE}_i$ ordered based on MSP is approximated} as
{
\begin{align}
\mathbb{P} (\tilde{C}_i)\!\approx\!\!\! \int\limits_0^{\infty} \! \int\limits_0^{x/2} \!\! e^{-{\theta_i r^{\eta} \sigma^2}} \!\! \mathcal{L}_{I^{\rm\o}_i \mid R_i,\rho } \!\left( {\theta_i r^{\eta}} \right) \! f_{R_i\mid \rho}(\!r\!\! \mid\!\! x \!)  dr f_{\rho}(\!x\!) dx  \label{sinr_OMA},
\end{align}
where $f_{\rho}(x)$ is given in \eqref{f_rho}, {$f_{R_i\mid \rho}(r \mid x)$ in \eqref{f_Ri} for Models 1 and 2 and \eqref{f_Ri_Mod4} for Model 3, and $\mathcal{L}_{I^{\rm\o}_i \mid R_i,\rho }$ is approximated in Lemmas 1, 2, and 3 for Models 1, 2, and 3, respectively.}
\begin{IEEEproof}
Using the exponential distribution of $h_i$ and the LT of $I^{\rm\o}_i$ conditioned on $R_i$ and $\rho$ we obtain \eqref{sinr_OMA}. 
\end{IEEEproof}
\textbf{\emph{Lemma 6:}} In the OMA network, the coverage probability of the typical $\text{UE}_i$ ordered based on IS$\oset[-.1ex]{_{\o}}{\raise.01ex\hbox{I}}$NR is approximated as
{
\begin{align}
\mathbb{P} (\tilde{C}_i) \approx \int_0^{\infty}  \left(1- F_{Z_i \mid \rho}(\theta_i \mid x) \right) f_{\rho}(x) dx \label{sinr_OMA_CB},
\end{align}
where $F_{Z_i \mid \rho}(y \mid \rho)$ is given in \eqref{F_Z} and $f_{\rho}(x)$} in \eqref{f_rho}.
\begin{IEEEproof}
\eqref{sinr_OMA_CB} follows from $\mathbb{P} (\tilde{C}_i)=\mathbb{P} \left( Z_i > \theta_i  \right)$.
\end{IEEEproof}
{Denote by $T_i$ the fraction of the time slot allotted to $\text{UE}_i$.} For a given SINR threshold $\theta_i$ and corresponding transmission (normalized) rate $\log(1+\theta_i)$, the throughput of {the typical} $\text{UE}_i$ is 
\begin{align}
\mathcal{R}_i= T_i \; \mathbb{P} (\tilde{C}_i) \log(1+\theta_i). \label{R_i_OMA}
\end{align}

\section{NOMA Optimization}
\subsection{Problem Formulation}\label{IVa}
Determining the optimization objective of the NOMA framework can be complicated. The objective of NOMA is to provide coverage to multiple UEs in the same time-frequency resource block. Naturally we are interested in maximizing the cell sum rate. It is well known that the cell sum rate is maximized by allocating all resources (power in the NOMA {network}) to the best UE \cite{N12}. However, this comes at the price of {a complete loss of} {fairness among NOMA UEs, which is one of the main motivations behind serving multiple UEs in a NOMA fashion}. Hence, we constrain the objective of maximizing cell sum rate to ensure multiple UEs are served. In addition to the power constraint we consider two kinds of constraints: 1) constraining resources so that each of the typical UEs achieves at least the {threshold minimum throughput (TMT)}, 2) constraining resources so that the typical UEs achieve symmetric (identical) throughput. Formally stated, these objectives are:
\begin{itemize}
\item $\mathcal{P}1$ - Maximum cell sum rate subject to the TMT $\mathcal{T}$:
\begin{align*}
&\max\limits_{ (P_i,\theta_i)_{{i=1},\ldots,N}}  \mathcal{R}_{\rm tot} \\
& \text{ subject to }  \sum\limits_{i=1}^N P_i= 1 \text{ and } \mathcal{R}_i \geq \text{$\mathcal{T}$}.
\end{align*}
Because this problem is non-convex, an optimum solution, {i.e., choice of $P_i$ and $\theta_i$ that result in the maximum constrained $\mathcal{R}_{\rm tot}$,} cannot be found using standard {optimization} methods. However, from the rate region for static channels we know that a RA that results in all UEs achieving the TMT $\mathcal{T}$, while all of the remaining power being allocated to the nearest UE, i.e., $\text{UE}_1$, to maximize its throughput is the optimum solution for that problem. An example of this for the two-user case is presented in \cite{N14}.
\item $\mathcal{P}2$ - Maximum symmetric throughput:
\begin{align*}
&\max\limits_{ (P_i,\theta_i)_{{i=1},\ldots,N}}  \mathcal{R}_{\rm tot}  \\
& \text{ subject to } \sum\limits_{i=1}^N P_i= 1 \text{ and } \mathcal{R}_1=\ldots= \mathcal{R}_N .
\end{align*}
This is equivalent to maximizing the smallest UE throughput. 
Solving this results in {a RA that achieves} the largest symmetric throughput (universal fairness), i.e., $\mathcal{R}_1=\ldots=\mathcal{R}_N$. Since this problem is also non-convex, an optimum solution cannot be found using standard {optimization} methods.
\end{itemize}
\textbf{\textit{Remark 4:}} The maximum symmetric throughput is the largest TMT that can be supported.\\
\textbf{\textit{Remark 5:}} Due to outage, the typical UEs that achieve the same {throughput ($\mathcal{R}_i$)} do not necessarily have the same individual transmission rates (and corresponding $\theta_i$'s). 

The same objectives hold for OMA {networks}. The constrained resource allocated to the UEs, however, is time for TDMA instead of power for NOMA, i.e., $\sum_i T_i= 1$. The OMA UEs enjoy full power in their transmissions. Optimization over transmission rate is done similarly to NOMA.


\subsection{Case Study: $N=2$}

\begin{figure}[h]
\begin{minipage}[htb]{0.98\linewidth}
\centering\includegraphics[width=0.75\columnwidth]{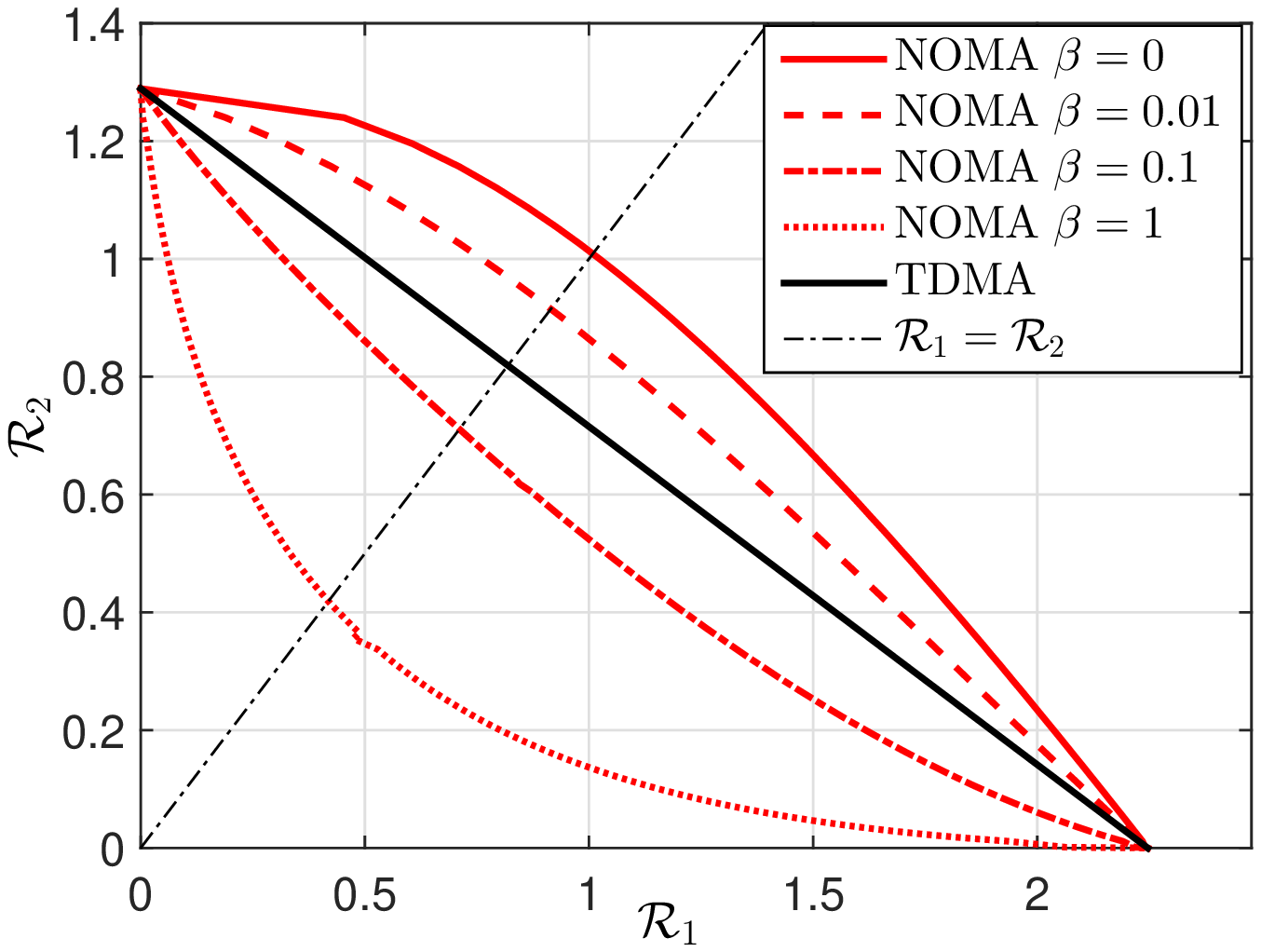}
\caption{Rate region for NOMA and TDMA with MSP-based UE ordering for Model 1 using different $\beta$ and $N=2$.\\}\label{betas_rateRegion}
\end{minipage}\;\;\;\;
\begin{minipage}[htb]{0.98\linewidth}
\centering\includegraphics[width=0.75\columnwidth]{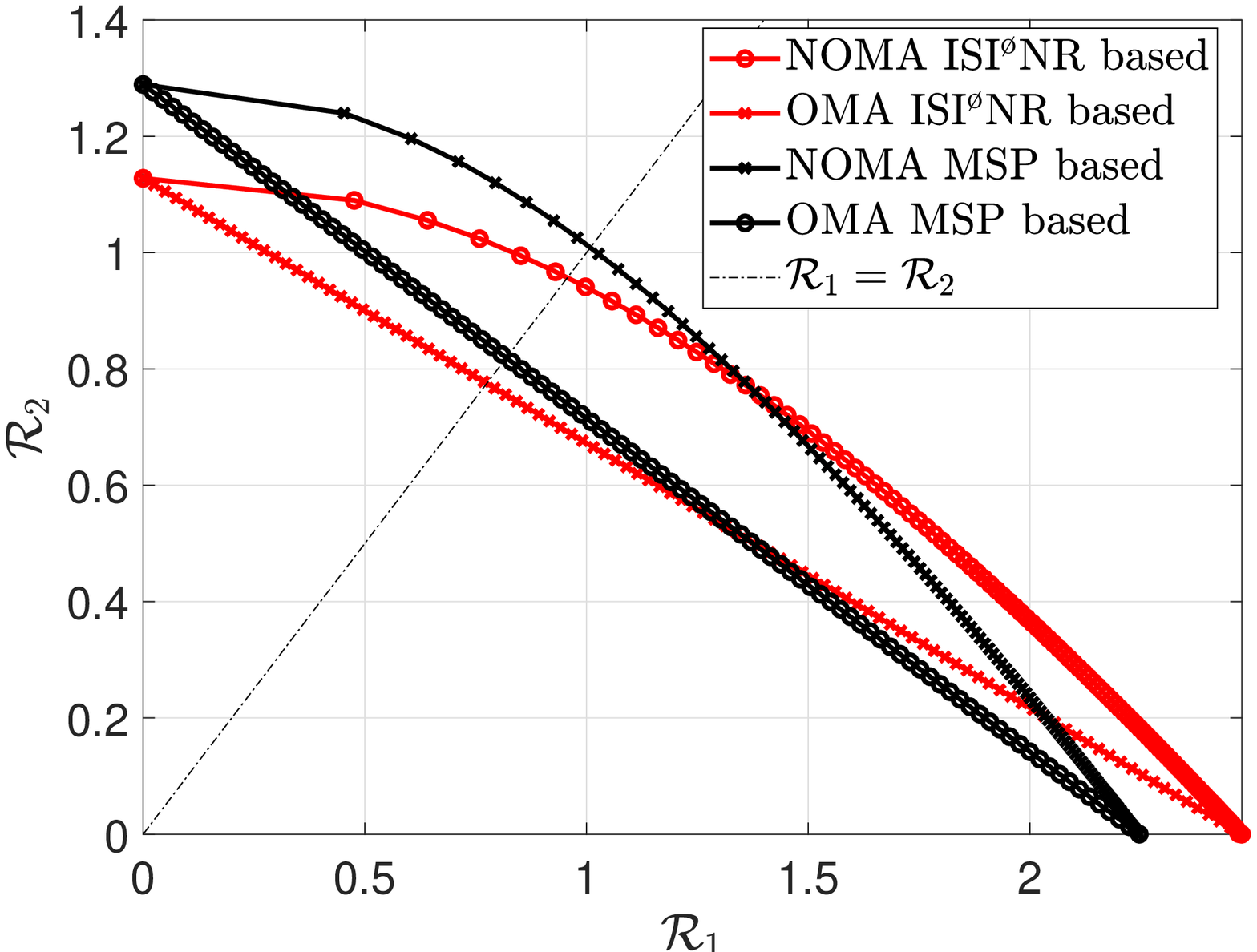}
\caption{Rate region for NOMA and TDMA with MSP and IS$\oset[-.1ex]{_{\o}}{\raise.01ex\hbox{I}}$NR-based UE ordering for Model 1 with $\beta=0$ and $N=2$.\\}\label{rate1vsrate2}
\end{minipage}\;\;\;\;
\begin{minipage}[htb]{0.98\linewidth}
\centering\includegraphics[width=0.75\columnwidth]{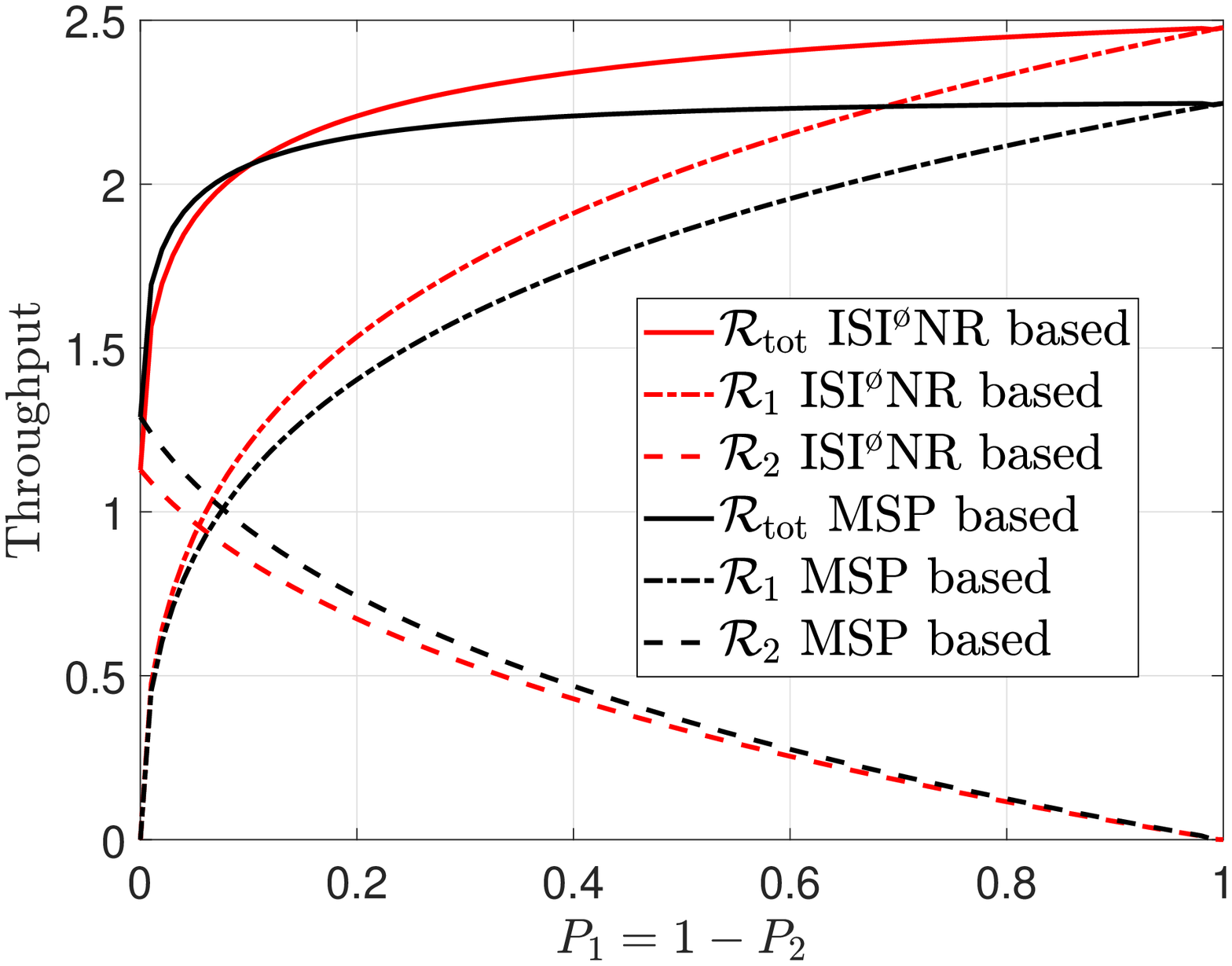}
\caption{Optimum cell sum rate and individual UE throughputs vs. $P_1$ for NOMA with MSP and IS$\oset[-.1ex]{_{\o}}{\raise.01ex\hbox{I}}$NR-based ordering for Model 1 with $\beta=0$ and $N=2$.}\label{Rtot_vsP}
\end{minipage}
\end{figure}



In this subsection we focus on the two-user case for which we can plot the maximum throughput for each UE subject to any power distribution for NOMA. This gives us the rate regions for the $N=2$ scenario as shown in Figs. \ref{betas_rateRegion} and \ref{rate1vsrate2}. We use {Model 1,} $\lambda=10$, $\sigma^2=-90$ dBm, $\eta=4$ in this subsection. {The rate regions in Fig. \ref{betas_rateRegion} are using different $\beta$ values and MSP-based ordering, while Fig. \ref{rate1vsrate2} uses both MSP- and IS$\oset[-.1ex]{_{\o}}{\raise.01ex\hbox{I}}$NR-based UE ordering with perfect SIC (i.e., $\beta=0$). Since the OMA scheme employed is TDMA, the RA in this case is not in terms of power but of time.} We use the optimal $\theta_i$ for a given power (time) distribution between the two NOMA (TDMA) UEs. 

{In the rate regions in Figs. \ref{betas_rateRegion}} and \ref{rate1vsrate2} {each point on the curve is obtained from optimal transmission rate allocation that maximizes throughput given a power (time) distribution for the two NOMA (TDMA) UEs.} {A} zero throughput of $\text{UE}_1$ (the intersection of the curves with the y-axis) corresponds to all the power being allocated to $\text{UE}_2$ in the case of NOMA and all the time being allotted to $\text{UE}_2$ in the case of TDMA and vice versa for zero throughput of $\text{UE}_2$ (the intersection of the curves with the x-axis). The rest of the points in each NOMA curve (TDMA curve) are made of all possible power (time) distributions between the two UEs. {Each curve is the boundary of the corresponding rate region. Optimal RA allows us to operate on the boundary of the rate region.} This sort of graph also reveals what areas of throughput operation result in higher cell sum rate given a TMT constraint on the UEs. {Additionally, if a symmetric throughput is required, the rate region shows us the maximum throughput possible.} Obtaining the rate region for larger $N$, however, is impractical as it requires exhaustively going through the $2N-1$ parameters for RA. 

With perfect SIC (i.e., $\beta=0$), if RA is optimum, i.e., if we operate at the boundary of the rate region, NOMA outperforms TDMA for both the symmetric-throughput objective ($\mathcal{P}2$) and given any TMT ($\mathcal{P}1$) as shown in Figs. \ref{betas_rateRegion} and \ref{rate1vsrate2}. {In Fig. \ref{betas_rateRegion} we observe that increasing $\beta$ deteriorates performance by pushing the boundary of the rate region inward. Also, if $\beta$ is too high, with optimum RA, TDMA always outperforms NOMA. {Additionally, the rate region graphs} shed light on the importance of RA; suboptimum RA can result in significant deterioration in performance as one could be operating inside the rate region far from the boundary. Thus, appropriate RA is very important to fully exploit the potential of NOMA.} 


In Fig. \ref{Rtot_vsP} we plot the optimum cell sum rate and individual UE throughput for $N=2$ against increasing $P_1$ (decreasing $P_2$) for NOMA with the two UE ordering techniques. Intuitively, UE ordering that incorporates more information about the channel is more accurate and should result in superior performance given any power distribution. Accordingly, one may anticipate that IS$\oset[-.1ex]{_{\o}}{\raise.01ex\hbox{I}}$NR-based ordering, which takes into account path loss, fading, intercell interference and noise, to always be superior in terms of $\mathcal{R}_{\rm tot}$ to MSP-based ordering, which only accounts for path loss. Contrary to this expectation, we observe that IS$\oset[-.1ex]{_{\o}}{\raise.01ex\hbox{I}}$NR-based ordering is not always superior. In particular, below a certain $P_1$, MSP-based ordering outperforms IS$\oset[-.1ex]{_{\o}}{\raise.01ex\hbox{I}}$NR-based ordering in terms of cell sum rate. IS$\oset[-.1ex]{_{\o}}{\raise.01ex\hbox{I}}$NR-based ordering exceeds in performance when $P_1$ is increased beyond this. In Fig. \ref{rate1vsrate2} we observe that this holds for TDMA, too. {This occurs because:
\begin{enumerate}
\item IS$\oset[-.1ex]{_{\o}}{\raise.01ex\hbox{I}}$NR-based ordering does, in fact, incorporate more information about the channel; the weakest (strongest) UE in this case on average is weaker (stronger) than the weakest (strongest) UE in MSP-based ordering. This can be seen in Fig. \ref{Rtot_vsP} for $N=2$ where the weak (strong) UE of IS$\oset[-.1ex]{_{\o}}{\raise.01ex\hbox{I}}$NR-based ordering consistently underperforms (outperforms) its MSP counterpart. {This applies to both NOMA and OMA as it depends on the UE ordering.}
\item Additionally for NOMA, which employs SIC, $\text{UE}_N$ is unable to cancel SI for the messages of any other UE and therefore suffers the largest intraference. In the case of IS$\oset[-.1ex]{_{\o}}{\raise.01ex\hbox{I}}$NR-based ordering, unlike its MSP counterpart, $\text{UE}_N$ may not necessarily be the farthest UE from the BS making the impact of intraference larger; this further deteriorates the SINR and therefore the throughput of the IS$\oset[-.1ex]{_{\o}}{\raise.01ex\hbox{I}}$NR-based $\text{UE}_N$. 
\end{enumerate}
Hence, when $P_1$ is small in Fig. \ref{Rtot_vsP}, the impact on $\mathcal{R}_{\rm tot}$ of the larger $\mathcal{R}_2$ in MSP-based ordering is more significant than the impact on $\mathcal{R}_{\rm tot}$ of the larger $\mathcal{R}_1$ in IS$\oset[-.1ex]{_{\o}}{\raise.01ex\hbox{I}}$NR-based ordering. When $P_1$ increases the impact of the significance is reversed.}



From Fig. \ref{rate1vsrate2} we observe that for higher TMT values (including the symmetric throughput), MSP-based ordering outperforms IS$\oset[-.1ex]{_{\o}}{\raise.01ex\hbox{I}}$NR-based ordering in terms of $\mathcal{R}_{\rm tot}$. This will become obvious in the results section as well.

\subsection{Algorithm for Solving $\mathcal{P}1$}\label{IVb}
Since standard {optimization} techniques cannot be employed for {any} $N$, the optimum solution to $\mathcal{P}1$ can only be found exhaustively by searching over all combinations of power and transmission rate for each of the $N$ NOMA UEs. This, however, is an extremely tedious approach, particularly as $N$ increases. In this subsection we propose an efficient algorithm that, while not guaranteeing an optimum solution, finds a feasible solution, i.e., a solution that satisfies the constraints (but there is no guarantee that the cell sum rate is close to the global maximum).


{Given a certain power, $\text{UE}_1$ is able to achieve a larger throughput from this resource than any other UE.} It therefore makes sense to solve $\mathcal{P}1$ by first ensuring that all UEs other than the strongest achieve TMT with the smallest powers possible. This will leave the largest $P_1$ for $\text{UE}_1$. $\text{UE}_1$ can then maximize the cell sum rate by maximizing $\mathcal{R}_1$ with this power by finding the appropriate transmission rate. {In other words, our algorithm for $\mathcal{P}1$ solves
\begin{align*}
&\max\limits_{ (P_i,\theta_i)_{{i=1},\ldots,N}} \mathcal{R}_{1} \\
& \text{ subject to }  \sum\limits_{i=1}^N P_i= 1 \text{ and } \mathcal{R}_j = \text{$\mathcal{T}$}, \;\; 2\leq j \leq N.
\end{align*}
}
We tackle this problem by decoupling the choice of power and transmission rate; our algorithm finds the minimum possible power and corresponding smallest transmission rate\footnote{For an $i\in \{1,\ldots, N\}$, the function $\mathcal{R}_i(\theta_i)$ is monotonically increasing from zero and then monotonically decreasing to zero, with a unique maximum at a finite $\theta_i>0$. This is because for small $\theta_i$, $\mathbb{P}(C_i)$ is close to 1, hence $\mathcal{R}_i$ increases linearly with $\log(1+\theta_i)$, while for large $\theta_i$, $\mathbb{P}(C_i)$ goes to zero more quickly than $\log(1+\theta_i)$ grows. Hence, each $\mathcal{R}_i$ (except the maximum) can be satisfied by two $\theta_i$ values. We select the smaller value since it increases the coverage probability for all UEs that are to decode the $i^{\rm th}$ message.} that achieve $\mathcal{T}$ for $\text{UE}_2$ to $\text{UE}_N$ and allocates the remaining power to $\text{UE}_1$. $\text{UE}_1$ then optimizes its transmission rate (and therefore $\theta_1$) with the remaining power to maximize its throughput. If a UE cannot attain $\mathcal{T}$, the available power is insufficient and the algorithm is unable to find a feasible solution as the cluster size $N$ is too large to attain this TMT for all UEs. This can be remedied by either decreasing $\mathcal{T}$ or $N$. {Formally, we state the working of the algorithm in Algorithm \ref{algo1}.}

{\setstretch{1.0}
\begin{algorithm*}[htb]
\caption{RA of a feasible solution to $\mathcal{P}1$}
\begin{multicols}{2}
\begin{algorithmic}[1]
\STATE Begin with $\text{UE}_N$, $i=N$, $P=[\;\;]$, ${\theta}=[\;\;]$, $\mathcal{R}=[\;\;]$
\WHILE{ $i>0$} \label{step2}
\IF{ $i>1$} 
\FOR{ {$P_i=0:\Delta_P:1-\sum_{k=i+1}^N P_k$}  }
\FOR{ {$\theta_i=\theta_{LB}:\Delta_{\theta}:\theta_{UB}$}}
\STATE Calculate $\mathcal{R}_i$ using \eqref{R_i_NOMA} with \eqref{sinr_2} for MSP-based (with \eqref{covg_CB} for IS$\oset[-.1ex]{_{\o}}{\raise.01ex\hbox{I}}$NR-based) UE ordering
\IF{$\mathcal{R}_i \geq \mathcal{T}$ }
\STATE Update: $P=[P_i; \; P]$; $\theta=[\theta_i; \; \theta]$; $\mathcal{R}=[\mathcal{R}_i; \; \mathcal{R}]$; $i=i-1$
\STATE Go to \ref{step2}
\ENDIF
\ENDFOR
\IF{$P_i = 1-\sum_{k=i+1}^N P_k$}
\STATE TMT cannot be met for all UEs; \textbf{exit}
\ENDIF
\ENDFOR
\ELSE 
\STATE $P_1=1-\sum_{k=2}^N P_k $
\IF{  $P_1>0$}
\STATE {$\mathcal{R}_1^{\rm vec}=[\;\;]$}
\FOR{$\theta_1 > 0$}
\STATE Calculate $\mathcal{R}_1$ using \eqref{R_i_NOMA} with \eqref{sinr_2} for MSP-based (with \eqref{covg_CB} for IS$\oset[-.1ex]{_{\o}}{\raise.01ex\hbox{I}}$NR-based) UE ordering
\STATE Update $\mathcal{R}_1^{\rm vec}=[\mathcal{R}_1^{\rm vec} ; \mathcal{R}_1]$
\ENDFOR
\STATE Update: $\mathcal{R}_1=\max(\mathcal{R}_1^{\rm vec}) $ and corresponding $\theta_1$
\IF{ $\mathcal{R}_1 \geq$ $\mathcal{T}$}
\STATE  Update: $P=[P_1; \; P]$; $\theta=[\theta_1; \; \theta]$; $\mathcal{R}=[\mathcal{R}_1; \; \mathcal{R}]$; $i=0$
\STATE Go to \ref{step2}
\ELSE 
\STATE  TMT cannot be met for all UEs; \textbf{exit}\\
\ENDIF
\ELSE
\STATE TMT cannot be met for all UEs; \textbf{exit}\\
\ENDIF 
\ENDIF 
\ENDWHILE
\end{algorithmic}\label{algo1}
\end{multicols}
\end{algorithm*}
}

{Since SIC requires knowledge of RA for the weaker UEs in the decoding chain, we start with $\text{UE}_N$ in line 1. {Note that the range for transmission rate is $\theta_i \geq 0$; to make our search finite, our algorithm searches in the range $\theta_{LB} \leq \theta_i \leq \theta_{UB}$.} In lines 3 to 16, starting with zero power, we search for the smallest corresponding transmission rate, {starting with {$\theta_{LB}$} and increasing in steps of $\Delta_{\theta}$,} that can meet the TMT constraint. For a given $P_i$, {if no $\theta_i$ is found until $\theta_{UB}$ that} achieves TMT, the power is increased {in steps of $\Delta_P$}. Once the minimum power that can meet the TMT is found, this power and the corresponding minimum transmission rate is saved. We then move on to the next weakest UE, using the stored power and transmission rate for the stronger UEs. This process is repeated for the $N-1^{th}$ strongest UE to $\text{UE}_2$. If the TMT cannot be met for $\text{UE}_i$, $i \in \{2, \ldots, N\}$, even when all of the remaining power $1-\sum_{k=i+1}^N P_k$ is allocated to it, the power budget is not sufficient for the current TMT and we exit the algorithm in line 13. Otherwise, the throughput achieved in line 7 is $\mathcal{R}_i = \mathcal{T}$. After the minimum required powers to achieve the TMT have been allocated to $\text{UE}_2, \ldots \text{UE}_{N}$, we use the remaining power in lines 17 to 34 to maximize $\mathcal{R}_1$ by finding the appropriate $\theta_1$. If the remaining power is 0 or if $\mathcal{R}_1<\mathcal{T}$ in lines 32 and 29, respectively, the TMT cannot be met for all UEs and we exit the algorithm.}


The same algorithm is employed for OMA, except that throughputs are calculated using  using \eqref{R_i_OMA} with \eqref{sinr_OMA} for MSP-based (with \eqref{sinr_OMA_CB} for IS$\oset[-.1ex]{_{\o}}{\raise.01ex\hbox{I}}$NR-based) UE ordering, and the contending resource is $T$ instead of $P$. Note that since our problem includes intercell interference, our RA is intercell interference-aware.

{We} compared the solutions of Algorithm 1 with those found using an exhaustive search for the case of $N=2$ and different values of $\mathcal{T}$. It turns out that {for $N=2$} our solutions are, in fact, optimum. It is of course not possible to compare the results of Algorithm 1 with those from an exhaustive search for larger $N$ as it is computationally too expensive.
\subsection{Algorithm for Solving $\mathcal{P}2$}\label{IVc}
Since $\mathcal{P}2$, like $\mathcal{P}1$, is non-convex, the optimum solution cannot be found using standard {optimization} techniques. As mentioned in the previous subsection, doing an exhaustive search over all combinations of power and transmission rate for each of the $N$ NOMA UEs is extremely tedious. We propose an algorithm which, while not guaranteeing an optimum solution, finds a feasible solution. {Denote by $\mu$ the threshold throughput that each UE must achieve. Then our algorithm for $\mathcal{P}2$ solves the following:
\begin{align*}
&\max\limits_{ (P_i,\theta_i)_{{i=1},\ldots,N}}  \mu \\
& \text{ subject to }  \sum\limits_{i=1}^N P_i= 1 \text{ and } \mathcal{R}_i = \mu.
\end{align*}
}
As with Algorithm \ref{algo1}, starting with $\text{UE}_N$, we aim to find the smallest power that can achieve $\mu$. Unlike Algorithm \ref{algo1} which does this upto $\text{UE}_2$ only, in this case we do it until $\text{UE}_1$, i.e., for all the UEs so that the UEs have symmetric throughput {$\mu$}. If the total power consumed is less than the power budget, we increase $\mu$. However, if each UE cannot achieve $\mu$, the threshold throughput is too high and needs to be reduced. In this way we update $\mu$ until the highest $\mu$ that can be achieved by all UEs while consuming the full power budget is found. Formally, the algorithm is stated in {Algorithm \ref{algo2}}. 

{\setstretch{1.0}
\begin{algorithm*}[htb]
\caption{RA of a feasible solution to $\mathcal{P}2$}
\begin{multicols}{2}
\begin{algorithmic}[1] 
\STATE Begin with $\mathcal{\mu}=0.3$, $\mu_H=\infty$, $\mu_{\rm L}=0$, $\zeta=0$, $a=1.3$, $n=1$
\WHILE{$n$}
\IF{$\zeta=0$}
\IF{$\mu_{\rm H}=\infty$}
\STATE{$\mu=a\mu$}
\ELSE
\STATE{$\mu=\frac{\mu_{\rm H}+\mu}{2}$}
\ENDIF
\ELSE
\STATE{$\mu=\frac{\mu_{\rm L}+\mu}{2}$}
\ENDIF
\STATE{Begin with $\text{UE}_N$, $i=N$, $P=[\;\;]$, ${\theta}=[\;\;]$}
\WHILE{ $i>0$} \label{step2_2}
\FOR{ {$P_i=0:\Delta_P:1-\sum_{k=i+1}^N P_k$}  }
\FOR{ {$\theta_i=\theta_{LB}:\Delta_{\theta}:\theta_{UB}$}}
\STATE Calculate $\mathcal{R}_i$ using \eqref{R_i_NOMA} with \eqref{sinr_2} for MSP-based (with \eqref{covg_CB} for IS$\oset[-.1ex]{_{\o}}{\raise.01ex\hbox{I}}$NR-based) UE ordering
\IF{$\mathcal{R}_i \geq \mu$ }
\STATE Update: $P=[P_i; \; P]$; $\theta=[\theta_i; \; \theta]$; $\zeta=0$; $i=i-1$
\STATE Go to \ref{step2_2}
\ENDIF
\ENDFOR
\IF{$P_i = 1-\sum_{k=i+1}^N P_k$}
\STATE $\mu$ cannot be met for all UEs; update: $\zeta=1$
\STATE Go to \eqref{step2_4}
\ENDIF
\ENDFOR
\ENDWHILE 
\IF{$\zeta=1$} \label{step2_4}
\STATE{$\mu_{\rm H}=\mu$} 
\ELSE
\STATE{$\mu_{\rm L}=\mu$} 
\ENDIF
\IF{$\mu_{\rm H}-\mu_{\rm L}<0.01 \mu_{\rm H}$}
\STATE Algorithm has converged, update: $n=0$
\ENDIF
\ENDWHILE
\end{algorithmic}\label{algo2}
\end{multicols}
\end{algorithm*}
}
{In Algorithm \ref{algo2} all UEs must achieve the threshold throughput of $\mu$, which is executed in lines 12 to 27. {This is done by starting with $\text{UE}_N$ to find the smallest power and its corresponding smallest transmission rate that can attain $\mu$; once found, these are stored. We then move on to the next weakest UE, using the stored power and transmission rate for the stronger UEs. This process is repeated until $\text{UE}_1$. If there isn't sufficient power for a UE to attain $\mu$, the flag $\zeta$ in line 23 is updated from 0 to 1 denoting that the current threshold throughput $\mu$ is too high and we exit the while loop to update $\mu$; otherwise, the flag $\zeta=0$.} We begin the algorithm assuming the last $\mu=0.3$ and $\zeta=0$. The upper bound on the threshold throughput (which not all of the UEs can attain at once), $\mu_{\rm H}$, is initially set to $\infty$ and the lower bound on the threshold throughput (which all of the UEs can attain), $\mu_{\rm L}$, is set to 0. We update $\mu_{\rm H}$ in line 29 when a smaller value of $\mu$ is found which all of the UEs fail to attain, i.e., when $\zeta=1$. Similarly, $\mu_{\rm L}$ is updated in line 31 when a larger value of $\mu$ is found which all of the UEs can attain, i.e., when $\zeta=0$. This way we iteratively update $\mu$ to be the average of the most updated upper and lower bounds in lines 7 and 10. When the difference between $\mu_{\rm H}$ and $\mu_{\rm L}$ is smaller than a certain value, such as 1\% in line 37, we assume the algorithm has converged. This way we are able to find the largest symmetric throughput. It should be noted that we use the coefficient $a$ such that $a>1$; this allows us to update $\mu$ when we do not have available a finite $\mu_{\rm H}$ in line 5. Also, note that although the algorithm begins with $\mu=0.3$ and $\zeta=0$, the choice of these parameters is arbitrary; even if $\mu=0.3$ is not achievable by all of the UEs, since $\mu_{\rm H}$ will be updated in the next iteration, the algorithm will not function incorrectly.    } 


The same algorithm is employed for OMA, except that throughputs are calculated using  using \eqref{R_i_OMA} with \eqref{sinr_OMA} for MSP-based (with \eqref{sinr_OMA_CB} for IS$\oset[-.1ex]{_{\o}}{\raise.01ex\hbox{I}}$NR-based) UE ordering, and the contending resource is $T$ instead of $P$. 

\section{Results}

{In this section we consider BS intensity $\lambda=10$, noise power $\sigma^2=-90$ dB and $\eta=4$.}

\subsection{Performance}

\begin{figure}
\begin{minipage}[htb]{0.98\linewidth}
\centering\includegraphics[width=0.75\columnwidth]{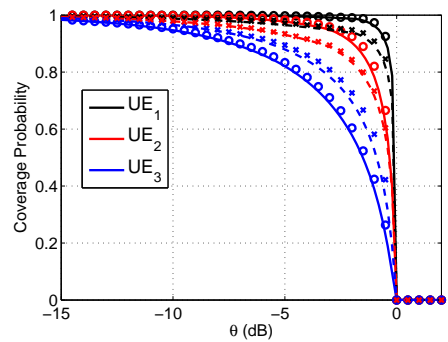}
\caption{SINR coverage vs. $\theta$ (identical transmission rate for all UEs) for Model 1 with $N=3$ employing the fixed PA $P_1=1/6$, $P_2=1/3$ and $P_3=1/2$. Solid (dashed) lines show the analysis for IS$\oset[-.1ex]{_{\o}}{\raise.01ex\hbox{I}}$NR-based (MSP-based) UE ordering. Markers show the Monte Carlo simulations.}\label{J_JointvsTheta_N3}
\end{minipage}
\end{figure}


We first show {using simulations} that the approximations in {Theorems 1 and 2} are tight. The results are shown in Fig. \ref{J_JointvsTheta_N3}, which considers a system with $N=3$ employing Model 1, a fixed PA scheme where $P_1=1/6$, $P_2=1/3$ and $P_3=1/2$, and both MSP-based and IS$\oset[-.1ex]{_{\o}}{\raise.01ex\hbox{I}}$NR-based UE ordering. For clarity of presentation we choose the same transmission rate for all three UEs in both cases and plot coverage of each UE against the corresponding SINR threshold. The figure verifies the accuracy of our SINR analysis as the coverage expressions for both types of UE ordering match the simulation closely. We observe that IS$\oset[-.1ex]{_{\o}}{\raise.01ex\hbox{I}}$NR-based UE ordering is superior for all UEs other than $\text{UE}_N$. As explained previously, this is because $\text{UE}_N$ for IS$\oset[-.1ex]{_{\o}}{\raise.01ex\hbox{I}}$NR-based ordering is weaker than its MSP counterpart. 

\begin{figure}
\begin{minipage}[htb]{0.98\linewidth}
\centering\includegraphics[width=0.75\columnwidth]{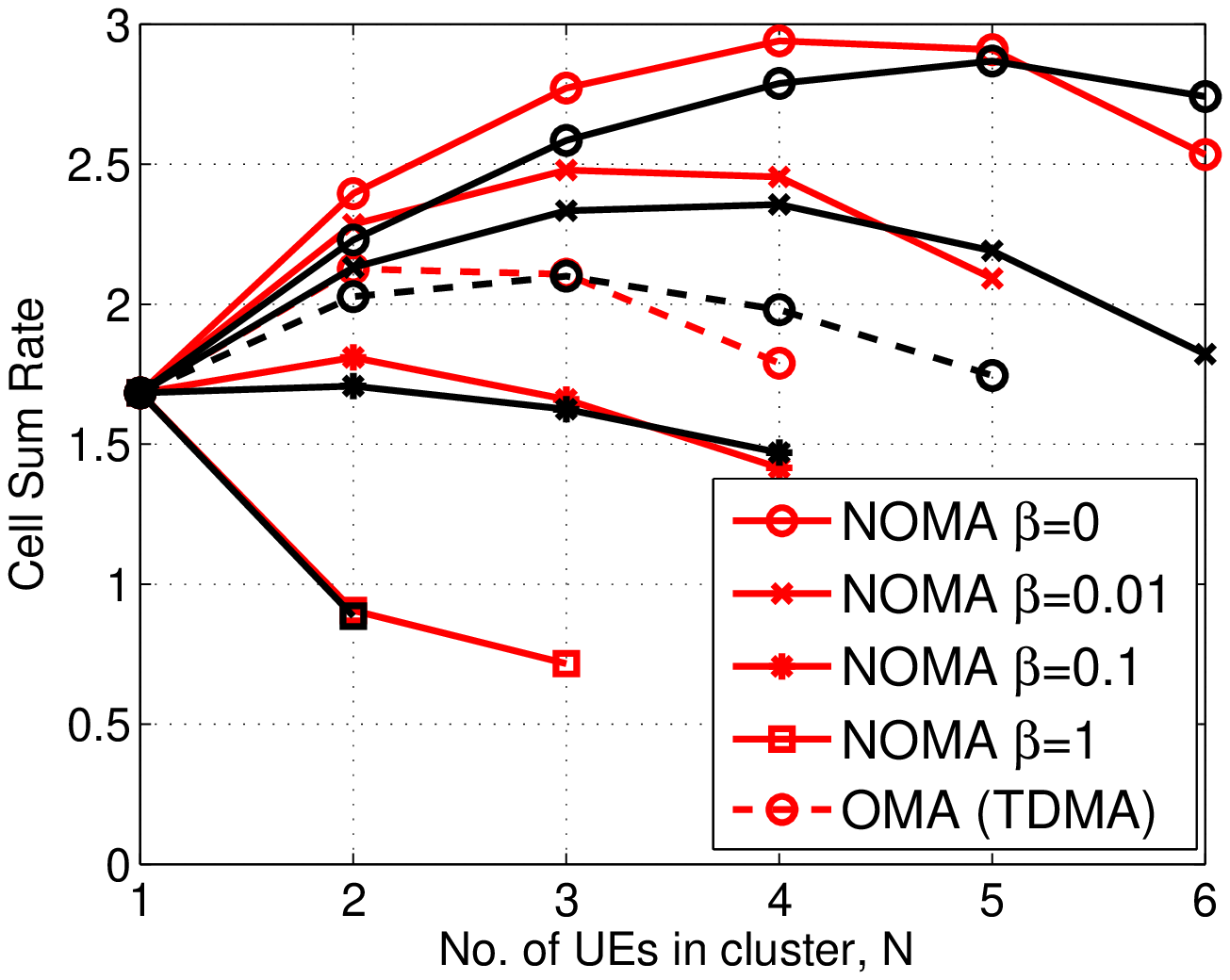}
\caption{$\mathcal{R}_{\rm tot}$ vs. $N$ for Model 1 using OMA and NOMA with $\mathcal{T}=0.3$ and different $\beta$ values. Black lines are for MSP-based UE ordering, while red lines are for IS$\oset[-.1ex]{_{\o}}{\raise.01ex\hbox{I}}$NR-based ordering. The curves end at the largest $N$ that can be supported given the TMT constraint and $\beta$.}\label{J_totRateVsN_QoS_betas}
\end{minipage}\;\;\;\;
\begin{minipage}[htb]{0.98\linewidth}
\centering\includegraphics[width=0.75\columnwidth]{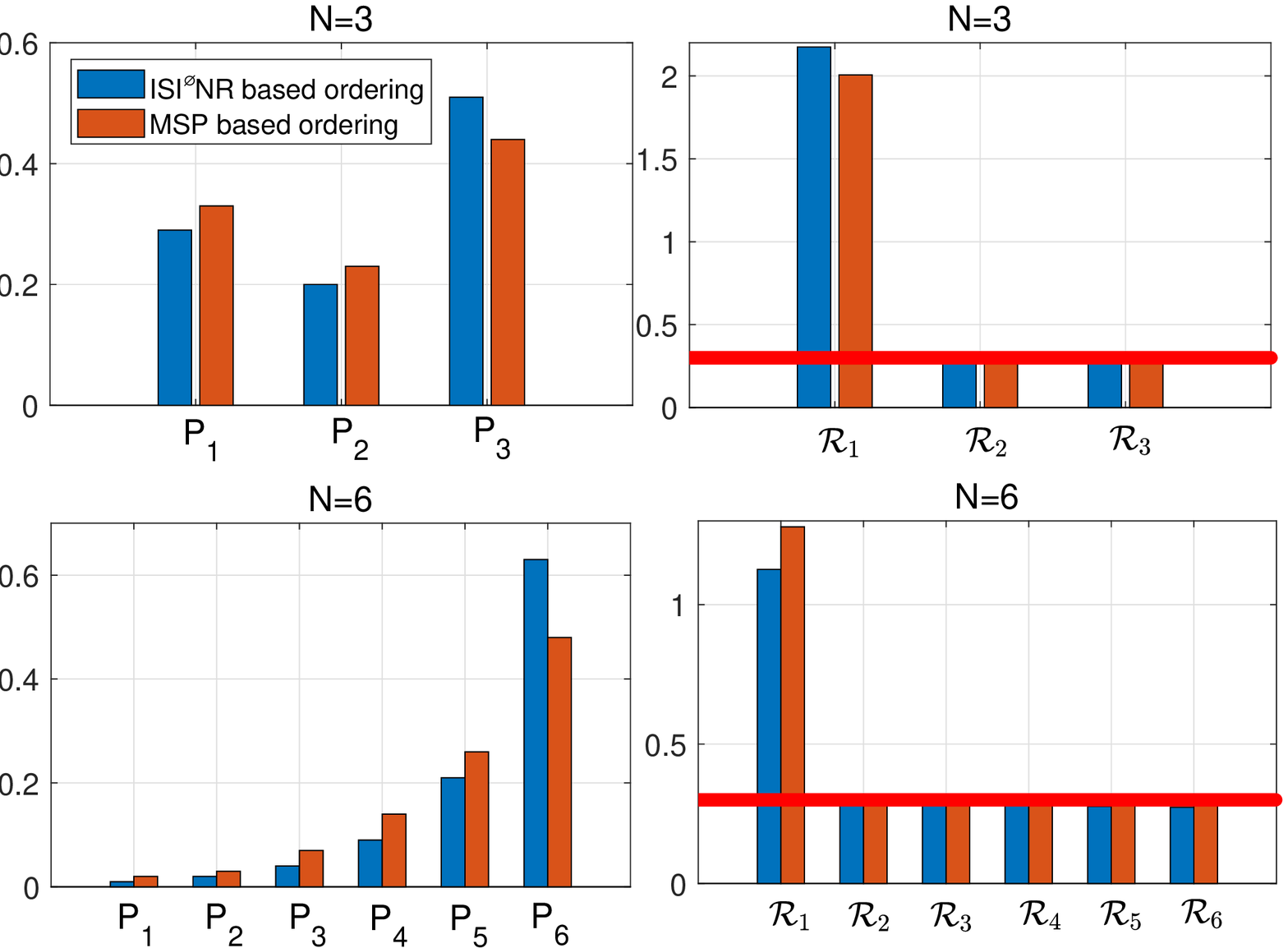}
\caption{Comparison of individual NOMA UE power and throughputs between IS$\oset[-.1ex]{_{\o}}{\raise.01ex\hbox{I}}$NR- and MSP-based UE ordering for $N=3$ and $N=6$ using Model 1 with $\beta=0$ and $\mathcal{T}=0.3$.\\}\label{J_detailedQOS}
\end{minipage}
\end{figure}

RA for the remaining figures is done according to Algorithms \ref{algo1} and \ref{algo2} depending on whether the objective is constrained to a TMT (i.e., $\mathcal{P}1$) or symmetric throughput (i.e., $\mathcal{P}2$), respectively. Unless specified otherwise, Model 1 is employed.

Fig. \ref{J_totRateVsN_QoS_betas} is a plot of the cell sum rate against the number of UEs in a NOMA cluster, $N$, for both MSP-based and IS$\oset[-.1ex]{_{\o}}{\raise.01ex\hbox{I}}$NR-based UE ordering given a TMT constraint. We have included $N=1$ in these plots which has the same $\mathcal{R}_{\rm tot}$ for all the curves since it only has one UE in a resource block ($\therefore$ independent of $\beta$) which maximizes its throughput ($\therefore$ independent of $\mathcal{T}$). Given $\mathcal{T}$ and $\beta$, there exists an optimum $N$ that maximizes $\mathcal{R}_{\rm tot}$. When $\beta$ is large we observe that using NOMA may not necessarily be more beneficial compared to OMA in terms of $\mathcal{R}_{\rm tot}$. Otherwise, for small $N$, increasing $N$ enhances $\mathcal{R}_{\rm tot}$ because interference cancellation is efficient in this regime, and more UEs are covered. {Also, increasing $N$ results in a stronger $\text{UE}_1$ as it decreases $R_1$ on average in the case of MSP-based ordering and improves $Z_1$ on average for IS$\oset[-.1ex]{_{\o}}{\raise.01ex\hbox{I}}$NR-based ordering; this enhances $\mathcal{R}_1$ given a $P_1$.} This in turn enhances $\mathcal{R}_{\rm tot}$ which in the TMT constraint problem ($\mathcal{P}1$) receives the largest contribution from $\mathcal{R}_1$. However, increasing $N$ beyond the optimum leaves too little power for $\text{UE}_1$ to boost $\mathcal{R}_1$ with. {For a given $\mathcal{T}$ and $\beta$, the resources are only sufficient to support a maximum cluster size; after this $N$ (discontinuation of the plots), not all of the UEs are able to achieve $\mathcal{T}$.} Increasing {$\beta$ results in a decrease in} the maximum cluster size that can be supported. {Similarly, increasing $\mathcal{T}$ results in a decrease in the maximum cluster size that can be supported \cite{myNOMA_icc}; this has not been shown for brevity.} NOMA outperforms OMA significantly if $\beta$ is small and can support the same number of UEs or more.


In Fig. \ref{J_totRateVsN_QoS_betas} we observe that for a given $\beta$ IS$\oset[-.1ex]{_{\o}}{\raise.01ex\hbox{I}}$NR-based UE ordering outperforms MSP-based ordering when $N$ is not larger than a certain value. After this, MSP performs better. In Fig. \ref{J_detailedQOS}, we compare the individual NOMA UE powers and throughputs when $\beta=0$ and $\mathcal{T}=0.3$ for $N=6$ (where MSP-based ordering outperforms) with $N=3$ (where IS$\oset[-.1ex]{_{\o}}{\raise.01ex\hbox{I}}$NR-based ordering outperforms). Unlike the other UEs, we observe that $\text{UE}_N$ requires more power to achieve $\mathcal{T}$ in the IS$\oset[-.1ex]{_{\o}}{\raise.01ex\hbox{I}}$NR-based ordering case than MSP-based ordering. This can be attributed to: 1) $\text{UE}_N$ in the IS$\oset[-.1ex]{_{\o}}{\raise.01ex\hbox{I}}$NR-based case is worse than $\text{UE}_N$ in the MSP-based case causing it to require more power, 2) $\text{UE}_N$ is unable to cancel any intraference; in IS$\oset[-.1ex]{_{\o}}{\raise.01ex\hbox{I}}$NR-based ordering $\text{UE}_N$ may not be the farthest UE, the impact of intraference may therefore be higher than its MSP-based counterpart where $\text{UE}_N$ is guaranteed to be farthest. The other non-strongest UEs in IS$\oset[-.1ex]{_{\o}}{\raise.01ex\hbox{I}}$NR-based ordering require lower powers to achieve TMT than their MSP counterparts implying they are on average stronger. For smaller $N$ ($N=3$ in Fig. \ref{J_detailedQOS}) when IS$\oset[-.1ex]{_{\o}}{\raise.01ex\hbox{I}}$NR-based ordering is employed, despite the increased power requirement by $\text{UE}_N$, there is still enough power left for $\text{UE}_1$ to maximize its throughput with so that $\mathcal{R}_{\rm tot}$ exceeds the MSP case. Although $\text{UE}_1$ in the IS$\oset[-.1ex]{_{\o}}{\raise.01ex\hbox{I}}$NR-based case is stronger, when $N$ is larger, {the $P_1$ left is too little and $\mathcal{R}_{\rm tot}$ is lower than its MSP-based counterpart.} The figures highlight that when $N$ is large, using the simpler MSP-based ordering scheme results in better performance than the complex IS$\oset[-.1ex]{_{\o}}{\raise.01ex\hbox{I}}$NR-based ordering.



\begin{figure}
\begin{minipage}[htb]{0.98\linewidth}
\centering\includegraphics[width=0.75\columnwidth]{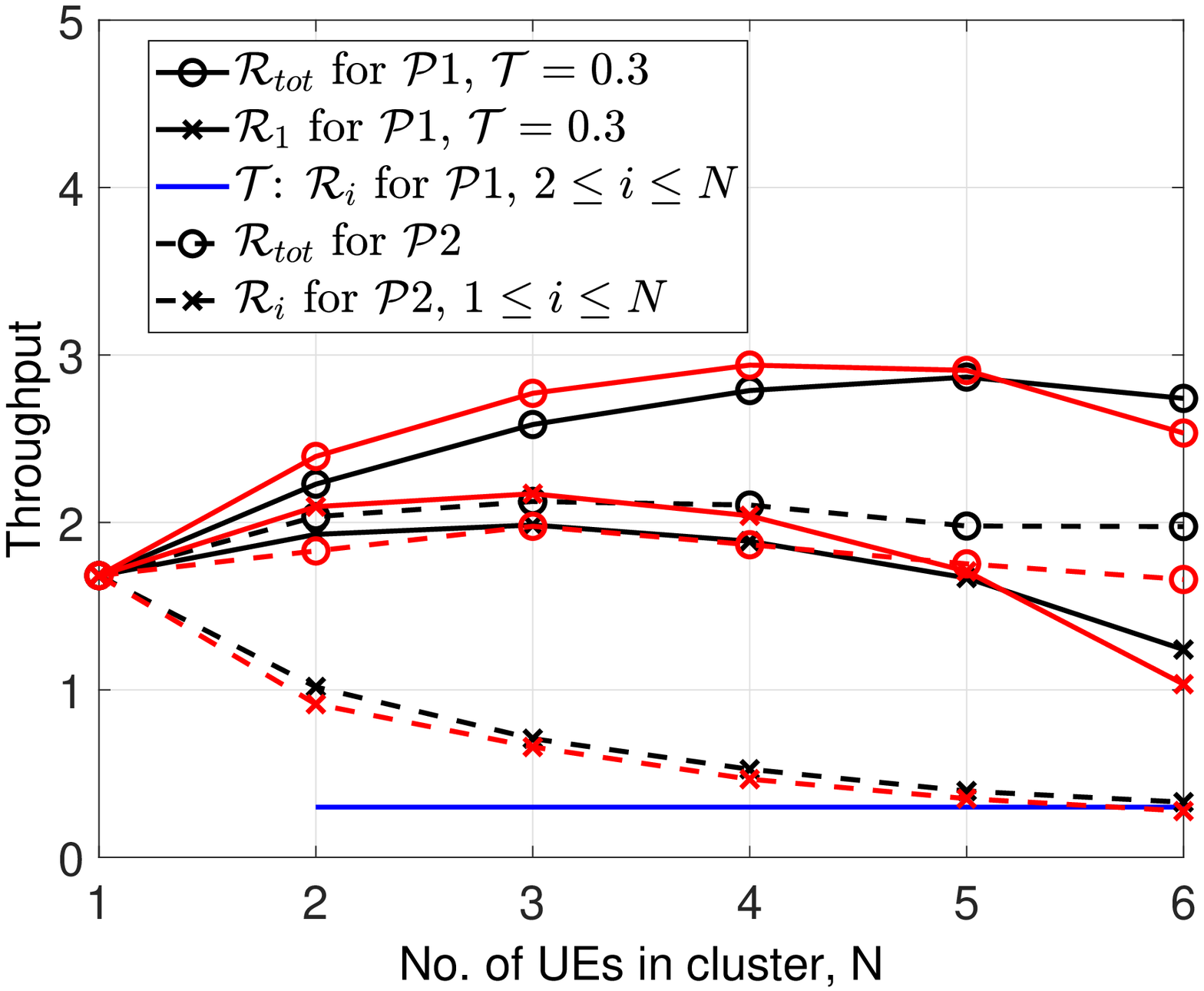}
\caption{Individual UE throughput and cell sum rate vs. $N$ using Model 1 with $\beta=0$ for: $\mathcal{P}1$ with $\mathcal{T}=0.3$ and $\mathcal{P}2$. Black lines are for MSP-based UE ordering, while red lines are for IS$\oset[-.1ex]{_{\o}}{\raise.01ex\hbox{I}}$NR-based. The blue line is the TMT achieved by $\text{UE}_2, \ldots, \text{UE}_N$ in $\mathcal{P}1$.}\label{J_totRate_qosVSsymm}
\end{minipage}\;\;\;\;
\begin{minipage}[htb]{0.98\linewidth}
\centering\includegraphics[width=0.75\columnwidth]{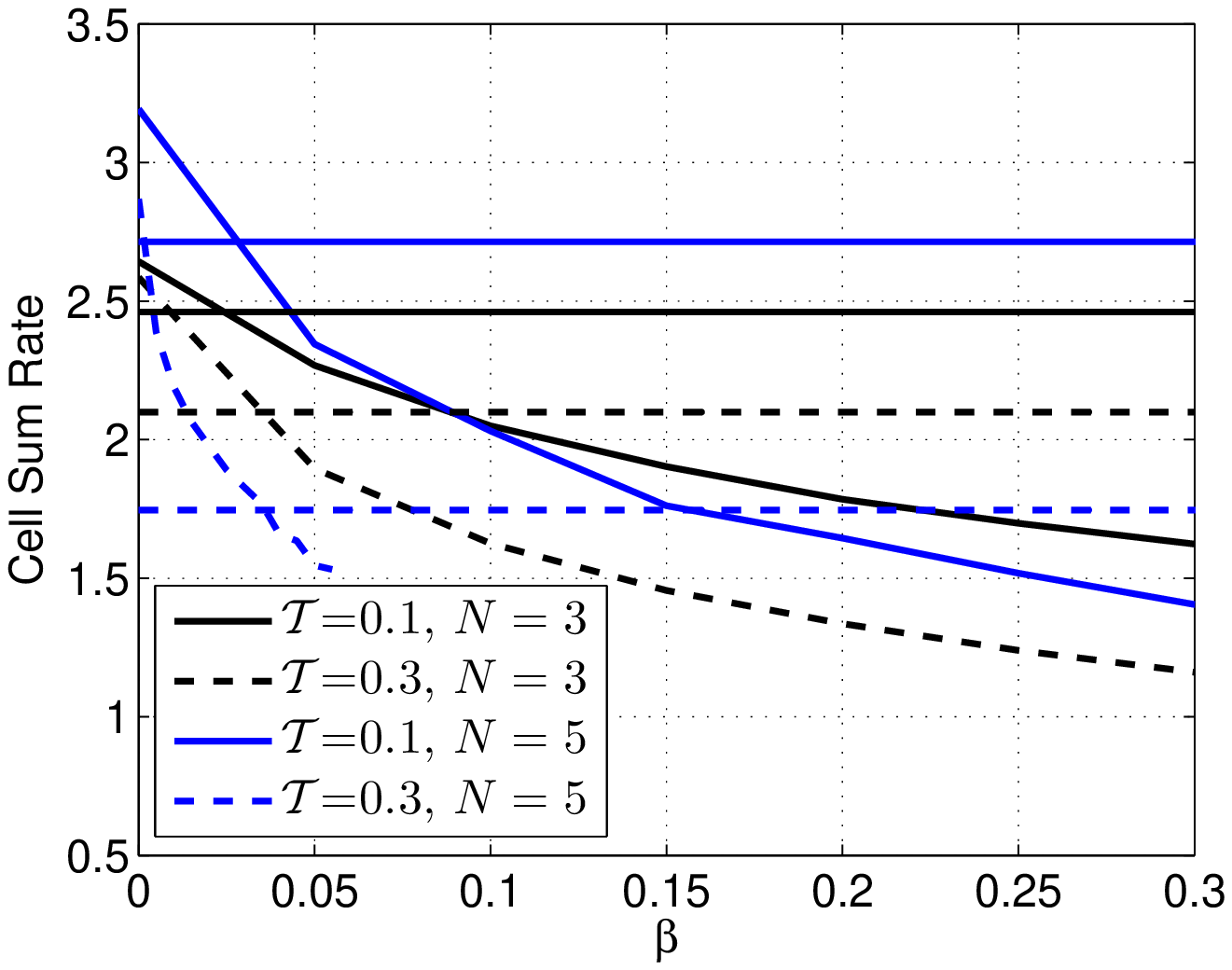}
\caption{$\mathcal{R}_{\rm tot}$ vs. $\beta$ for Model 1 with different values of $N$ and $\mathcal{T}$ using MSP-based ordering. Curves represent NOMA and horizontal lines (independent of $\beta$) represent OMA.\\}\label{vsBeta}
\end{minipage}
\end{figure}

Fig. \ref{J_totRate_qosVSsymm} is a plot of {both the $\mathcal{R}_{\rm tot}$ and individual UE throughput} against the number of UEs in a NOMA cluster for MSP and IS$\oset[-.1ex]{_{\o}}{\raise.01ex\hbox{I}}$NR-based UE ordering. We compare the maximum symmetric throughput objective in $\mathcal{P}2$ (dashed lines) with the TMT constrained objective where $\mathcal{T}=0.3$ in $\mathcal{P}1$ (solid lines). The curves for $\mathcal{P}1$ end at the largest $N$ that can be supported given the TMT constraint and $\beta$. {The symmetric throughput objective does not have a limit on the largest $N$ that can be supported but for comparison with $\mathcal{P}1$, we plot them up to the same value of $N$.} The TMT constraint of $\mathcal{T}=0.3$ always outperforms the symmetric throughput objective in terms of $\mathcal{R}_{\rm tot}$ for the values of $N$ that it can support; this is in accordance with what is anticipated from the rate region in Fig. \ref{rate1vsrate2}. Additionally, {for $\mathcal{P}2$}, MSP-based ordering always has a superior $\mathcal{R}_{\rm tot}$ compared to its IS$\oset[-.1ex]{_{\o}}{\raise.01ex\hbox{I}}$NR-based counterpart, which is in line with {our conclusions from Fig. \ref{rate1vsrate2}.} 

In Fig. \ref{J_totRate_qosVSsymm} the symmetric throughput objective of $\mathcal{P}2$, like the case of $\mathcal{P}1$, has an optimum $N$ that maximizes $\mathcal{R}_{\rm tot}$. For the problem in $\mathcal{P}2$, the largest symmetric throughput is limited by the weakest UE, $\text{UE}_N$, which requires the largest power. As $N$ grows, the weakest UE becomes worse and the total power needs to be shared among a larger number of UEs. This causes the individual UE's (symmetric) throughput to decrease with $N$ as shown. However, increasing $N$ at first enhances $\mathcal{R}_{\rm tot}$ because SIC is efficient in this regime and more UEs are covered, so the gains from the larger number of UEs are more significant. For larger $N$, the individual UE throughput becomes too small. Consequently $\mathcal{R}_{\rm tot}$ starts decreasing after the optimum $N$. {Additionally, as long as $N$ is not too large, the individual UEs perform better for $\mathcal{P}2$ compared to $\text{UE}_2, \ldots, \text{UE}_N$ in $\mathcal{P}1$ which achieve $\mathcal{T}$. Also, $\mathcal{R}_1$ in $\mathcal{P}1$ outperforms the individual UE throughput in $\mathcal{P}2$ as anticipated. More interestingly, $\mathcal{R}_1$ has an optimum $N$ for which it is maximized. This is due to the improving strength of $\text{UE}_1$ as $N$ grows, followed by a decrease in $\mathcal{R}_1$ because of lower available $P_1$ when $N$ is too large.}


Fig. \ref{vsBeta} plots the cell sum rate against $\beta$ for different $N$ and $\mathcal{T}$ using MSP-based ordering. Since OMA does not use SIC, the corresponding $\mathcal{R}_{\rm tot}$ plots are horizontal lines independent of $\beta$. The figure shows the existence of a maximum $\beta$ value until which a NOMA system with a particular $N$ and $\mathcal{T}$ is able to outperform the corresponding OMA system. {This highlights that there is a critical minimum level of SIC required for NOMA to outperform OMA.} We also observe that the decrease in $\mathcal{R}_{\rm tot}$ as a function of $\beta$ is steeper for larger $N$ and $\mathcal{T}$ highlighting their increased susceptibility to RI.  


{The results highlight the importance and impact of choosing network parameters such as $N$ and the UE ordering technique depending on the network objective and $\beta$. As an example, if complete user fairness is required, i.e., the objective is $\mathcal{P}2$, MSP-based ordering would result in higher $\mathcal{R}_{\rm tot}$, while $N$ would be chosen according to $\beta$. Similarly, if the network requires a certain TMT, the objective is $\mathcal{P}1$. To enhance $\mathcal{R}_{\rm tot}$, IS$\oset[-.1ex]{_{\o}}{\raise.01ex\hbox{I}}$NR-based ordering may be chosen if $\mathcal{T}$ is not too high with a smaller $N$; otherwise MSP-based ordering would be a better option. The value of $N$ would also depend on $\beta$.}

\begin{figure}[h]
\begin{minipage}[htb]{0.98\linewidth}
\centering\includegraphics[width=0.75\columnwidth]{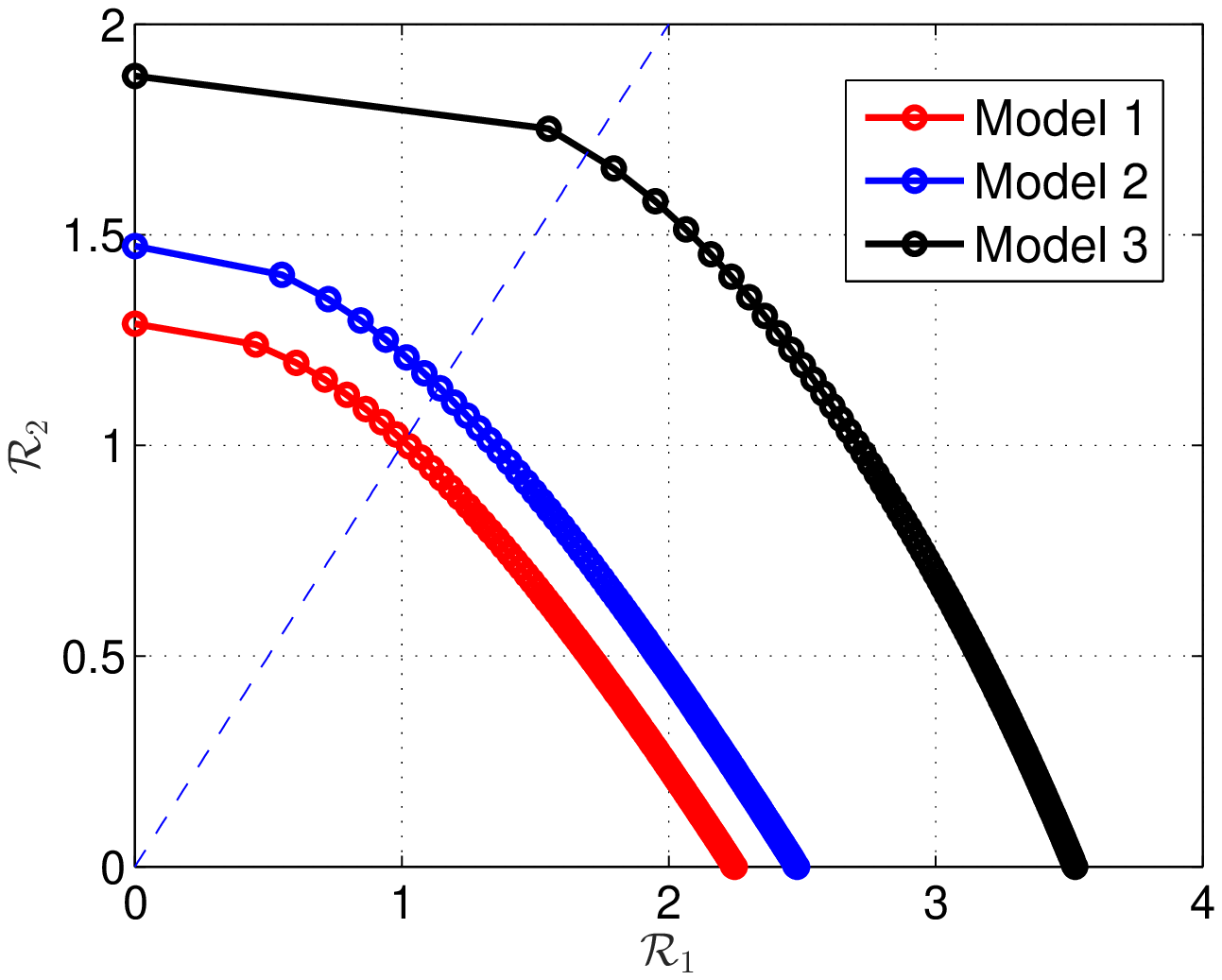}
\caption{Rate region for NOMA with MSP-based UE ordering, $\beta=0$ and $N=2$ for Models 1, 2, and 3.\\}\label{Mod123_rateRegion}
\end{minipage}\;\;\;\;
\begin{minipage}[htb]{0.98\linewidth}
\centering\includegraphics[width=0.75\columnwidth]{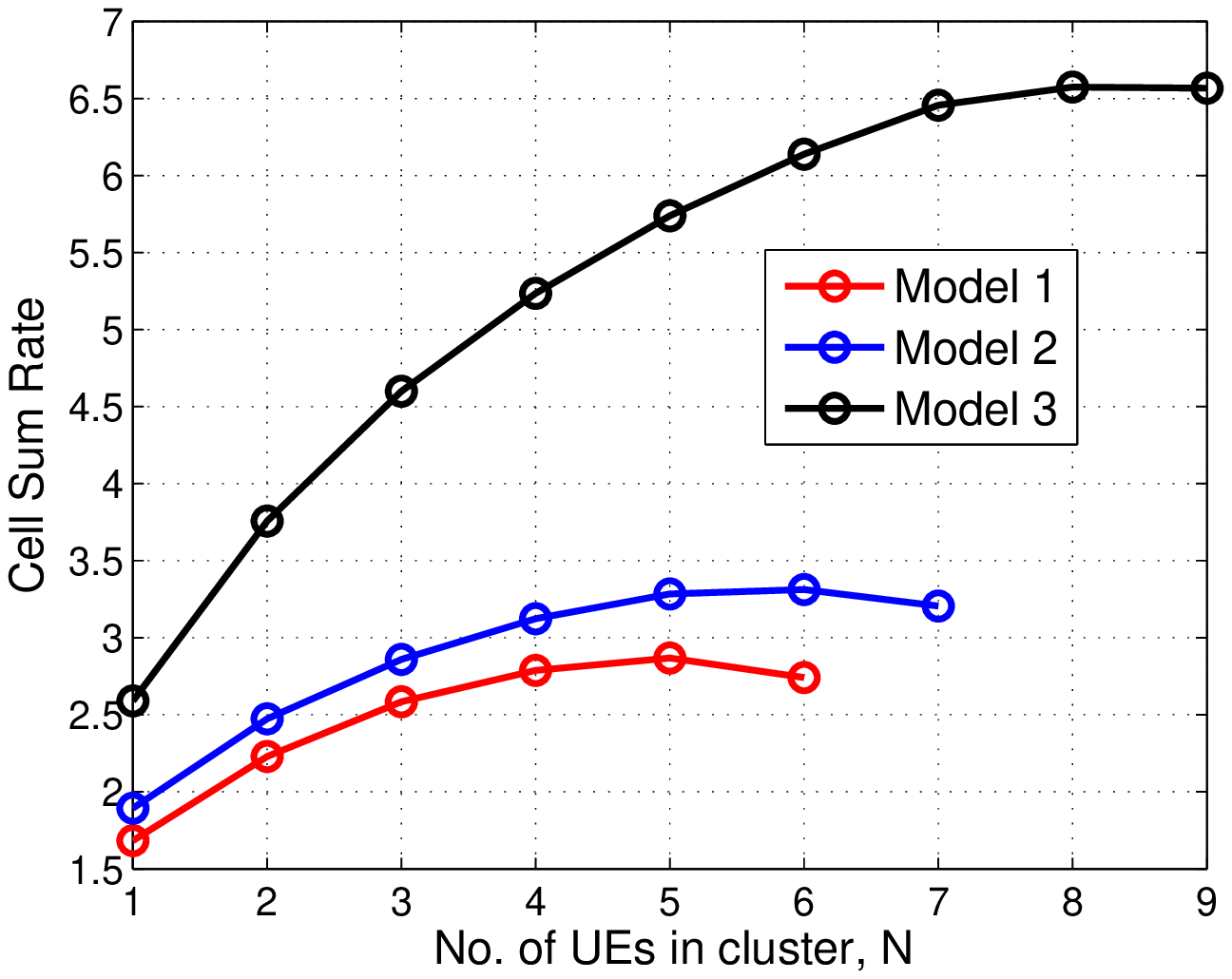}
\caption{$\mathcal{R}_{\rm tot}$ vs. $N$ for $\mathcal{T}=0.3$ using NOMA and $\beta=0$ for MSP-based ordering. The curves end at the largest $N$ that can be supported given $\mathcal{T}$ and $\beta$.}\label{Mod123_totRateVsN_QoS}
\end{minipage}
\end{figure}

Fig. \ref{Mod123_rateRegion} is a plot of the rate regions for the $N=2$ case using MSP-based UE ordering and with $\beta=0$ for Models 1, 2, and 3. We observe that the optimum performance (rate region boundary) of Model 2 can be viewed as an upper bound on that of Model 1. This can be explained by the fact that Model 2 {selects UEs that are located farther from the closest cell edge} than the UEs in Model 1 resulting in better average {interference} conditions and consequently, performance. Similarly Model 3, which {selects UEs that are located even farther from the closest cell edge} than both Models 1 and 2 and thereby have the best average {interference} conditions, upper bounds the other two models in terms of performance.

Fig. \ref{Mod123_totRateVsN_QoS} is a plot of the cell sum rate with increasing $N$ for all three models with MSP-based ordering, $\beta=0$ and $\mathcal{T}=0.3$. In general, {the smaller the sector angle $\phi$,} the more clustered NOMA UEs are towards the center of the cell and the better their performance is on average. Accordingly, we observe that Model 2 outperforms Model 1 for each value of $N$, and that Model 3 outperforms both Models 1 and 2. {This highlights how a superior UE clustering technique that selects UEs with good interference conditions is able to significantly improve the performance. In particular, for the given $\beta$ and $\mathcal{T}$ at its optimum $N$, Model 2 outperforms the optimum of Model 1 by 15.5\%, and Model 3 outperforms Models 1 and 2 by 129\% and 98.4\%, respectively.} Additionally, we observe that a more superior clustering technique is able to support a larger maximum cluster size given a TMT constraint; for $\mathcal{T}=0.3$, Models 2 and 3 are able to support a largest $N$ of 7 and 9, respectively, compared to the largest $N$ of 6 in the case of Model 1.

It ought to be highlighted that although Model 3 shows a significant improvement in performance, {its main purpose is to serve as an upper bound. In a practical setting, a model with a sector angle such as $\phi=\pi/2$ (i.e., ``Model 2.5'') may provide a very good trade off between having enough UEs available for NOMA and the interference conditions.}


\subsection{Complexity}
In this subsection we discuss the complexity of the proposed algorithms and compare them with an exhaustive search. {As mentioned in Section IV, since the range for transmission rate is $\theta_i \geq 0$, our algorithms search in $\theta_{LB} \leq \theta_i \leq \theta_{UB}$ to make the search finite. For a fair comparison we use the same search space for the exhaustive search. We search in} $-10 \text{ dB} \leq \theta_i \leq 22 \text{ dB}$ and use step size $\Delta_{\theta}=1$ in our work. As a result there are $\hat{\Delta}_{\theta}=33$ choices of $\theta_i$. Since the range of power allocated to a UE is $0\leq P_i \leq 1$, based on the step size $\Delta_P$ there are $1/\Delta_P$ choices of $P_i$. We define the complexity as the sum of the number of times throughput is calculated for each UE, i.e., the number of power-transmission throughput combinations the algorithm iterates over. Consequently, for an exhaustive search the complexity is $( \hat{\Delta}_{\theta}/\Delta_P)^{N}$.  

Fig.~\ref{cmplx}(\subref{cmplx1}) plots the complexity against $N$ for different values of TMT and step size $\Delta_P$ for Algorithm 1. As anticipated, decreasing $\Delta_P$ increases the complexity. Also, decreasing the TMT decreases complexity as the non-strongest UEs find the least power required to achieve the TMT more quickly. We observe that for a given $\mathcal{T}$ and $\Delta_P$, IS$\oset[-.1ex]{_{\o}}{\raise.01ex\hbox{I}}$NR-based ordering has lower complexity than its MSP-based counterpart until large $N$ where its complexity becomes larger. This is based on similar reasons to Fig. \ref{J_detailedQOS} where $\text{UE}_N$ in IS$\oset[-.1ex]{_{\o}}{\raise.01ex\hbox{I}}$NR-based ordering requires larger power than its MSP counterpart, which increases the complexity and becomes the dominant factor at high $N$. {The results suggest that the complexity is of the form $c^N$, where $c=\hat{\Delta}_{\theta}/\Delta_P$ for an exhaustive search. In the case of our algorithms we observe that $c \ll (\hat{\Delta}_{\theta}/\Delta_P)$. In particular, for Algorithm 1 with $\Delta_P=0.01$, our $c$ is about 3 and thus 1000 times smaller than $\hat{\Delta}_{\theta}/\Delta_P=3300$.}

In Fig.~\ref{cmplx}(\subref{cmplx2}) we additionally plot the complexity curves for an exhaustive search and Algorithm 2. Since Algorithm 2 has Algorithm 1 nested in it, it repeats Algorithm 1 multiple times in a way; consequently, the complexity is higher. Increasing the step size has a similar effect as in Algorithm 1, we do not show these for brevity. It should be noted that the complexity of Algorithm 2 does not increase monotonically with $N$ as in the case of Algorithm 1. This is due of the more heuristic nature of Algorithm 2 because of the choice of the initial parameters $a$ and $\mu$ which result in a varying number of iterations before the largest symmetric throughput is achieved. As a result, complexity can change depending on the choice of these parameters. For fairness, we chose the same parameters for all $N$ corresponding to the values mentioned in Algorithm \ref{algo2}. Most importantly, from Fig.~\ref{cmplx}(\subref{cmplx2}) we observe the significant difference between the complexity of our algorithms and an exhaustive search. Since an exhaustive search is the only optimum way for solving both non-convex problems $\mathcal{P}1$ and $\mathcal{P}2$, the stark difference in complexity motivates the use of efficient algorithms such as ours for RA. 



\begin{figure}
\begin{subfigure}[htb]{0.98\linewidth}
\centering\includegraphics[width=0.75\columnwidth]{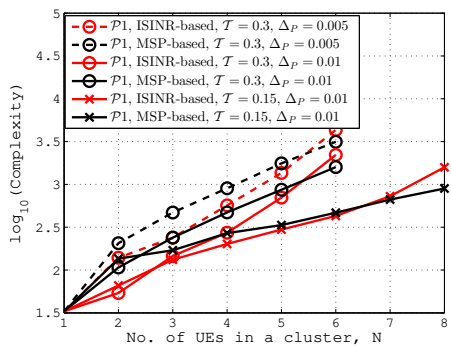}
\subcaption{Algorithm 1, for solving $\mathcal{P}1$, with $\beta=0$ for different $\mathcal{T}$ and $\Delta_P$.}\label{cmplx1}
\end{subfigure}\;\;\;\;
\begin{subfigure}[htb]{0.98\linewidth}
\centering\includegraphics[width=0.75\columnwidth]{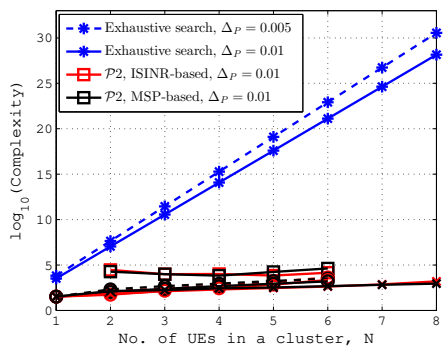}
\subcaption{{Algorithm 1 and Algorithm 2 with $\beta=0$, and exhaustive search for different $\Delta_P$.}}\label{cmplx2}
\end{subfigure}
\caption{Complexity vs. $N$ for NOMA UEs selected according to Model 1.}\label{cmplx}
\end{figure}

\section{Conclusion}\label{Conc}
In this paper a large cellular network that employs NOMA in the downlink is studied. As NOMA requires ordering of the UEs based on some measure of link quality, two kinds of UE ordering techniques are analyzed and compared: 1) MSP-based ordering, 2) IS$\oset[-.1ex]{_{\o}}{\raise.01ex\hbox{I}}$NR-based ordering. {An SINR analysis that takes into account the SIC chain and RI from imperfections in SIC is developed for each ordering technique.} We show that neither ordering technique is consistently superior to the other and highlight scenarios where each outperforms the other. Two non-convex problems of maximizing the cell sum rate $\mathcal{R}_{\rm tot}$ subject to a constraint are formulated: a TMT constraint $\mathcal{T}$ in $\mathcal{P}1$ and the symmetric throughput constraint in $\mathcal{P}2$. Since the optimum solution for RA to solve each problem requires an exhaustive search, two efficient algorithms for general NOMA cluster size $N$ that give feasible solutions for intercell interference-aware PA and choice of transmission rate are proposed. {We show that the complexity of the proposed algorithms is significantly lower than an exhaustive search.} Additionally, the existence of an optimum NOMA cluster size that maximizes $\mathcal{R}_{\rm tot}$ for each problem is shown. {It is observed} that $\mathcal{P}1$ provides a higher $\mathcal{R}_{\rm tot}$; however, $\mathcal{P}2$ guarantees better individual UE performance. {Furthermore, it is shown that} NOMA outperforms OMA provided $\beta$ is below a certain {critical} value. {The results highlight} the importance and impact of choosing network parameters such as $N$ and the UE ordering technique, depending on the network objective and $\beta$. Three models to show the impact of UE clustering in NOMA are introduced. The models demonstrate that efficient UE clustering which selects UEs with good interference conditions can improve performance significantly; {in fact, with efficient UE clustering the cell sum rate can be doubled.} 

\appendices

\bibliographystyle{IEEEtran}
\bibliography{refsNOMA}

\begin{thebibliography}{10}
\providecommand{\url}[1]{#1}
\csname url@samestyle\endcsname
\providecommand{\newblock}{\relax}
\providecommand{\bibinfo}[2]{#2}
\providecommand{\BIBentrySTDinterwordspacing}{\spaceskip=0pt\relax}
\providecommand{\BIBentryALTinterwordstretchfactor}{4}
\providecommand{\BIBentryALTinterwordspacing}{\spaceskip=\fontdimen2\font plus
\BIBentryALTinterwordstretchfactor\fontdimen3\font minus
  \fontdimen4\font\relax}
\providecommand{\BIBforeignlanguage}[2]{{%
\expandafter\ifx\csname l@#1\endcsname\relax
\typeout{** WARNING: IEEEtran.bst: No hyphenation pattern has been}%
\typeout{** loaded for the language `#1'. Using the pattern for}%
\typeout{** the default language instead.}%
\else
\language=\csname l@#1\endcsname
\fi
#2}}
\providecommand{\BIBdecl}{\relax}
\BIBdecl

\bibitem{myNOMA_icc}
K.~S. Ali, H.~Elsawy, A.~Chaaban, M.~Haenggi, and M.~S. Alouini, ``Analyzing
  non-orthogonal multiple access ({NOMA}) in downlink {P}oisson cellular
  networks,'' in \emph{Proc. of IEEE International Conference on Communications
  (ICC18)}, 2018.

\bibitem{N12_9}
T.~M. Cover and J.~A. Thomas, \emph{Elements of Information Theory}.\hskip 1em
  plus 0.5em minus 0.4em\relax NJ: John Wiley, 2006.

\bibitem{book_fundWirelessComm}
D.~Tse and P.~Viswanath, \emph{Fundamentals of Wireless Communication}.\hskip
  1em plus 0.5em minus 0.4em\relax Cambridge University Press, 2004.

\bibitem{N12_16}
P.~Patel and J.~Holtzman, ``Analysis of a simple successive interference
  cancellation scheme in a {DS/CDMA} system,'' \emph{IEEE J. Selec. Areas
  Commun.}, vol.~12, no.~5, pp. 796--807, June 1994.

\bibitem{sup1}
S.~Vanka, S.~Srinivasa, Z.~Gong, P.~Vizi, K.~Stamatiou, and M.~Haenggi,
  ``Superposition coding strategies: Design and experimental evaluation,''
  \emph{IEEE Trans. Wireless Commun.}, vol.~11, no.~7, pp. 2628--2639, July
  2012.

\bibitem{N1}
Y.~Saito \emph{et~al.}, ``Non-orthogonal multiple access ({NOMA}) for cellular
  future radio access,'' in \emph{Proc. of IEEE 77th Vehicular Technology
  Conference (VTC Spring 2013)}, June 2013, pp. 1--5.

\bibitem{N8}
Z.~Ding, Z.~Yang, P.~Fan, and H.~V. Poor, ``On the performance of
  non-orthogonal multiple access in {5G} systems with randomly deployed
  users,'' \emph{IEEE Signal Proc. Letters}, vol.~21, no.~12, pp. 1501--1505,
  Dec. 2014.

\bibitem{N9}
S.~Timotheou and I.~Krikidis, ``Fairness for non-orthogonal multiple access in
  {5G} systems,'' \emph{IEEE Signal Proc. Letters}, vol.~22, no.~10, pp.
  1647--1651, Oct. 2015.

\bibitem{N13}
J.~Choi, ``Power allocation for max-sum rate and max-min rate proportional
  fairness in {NOMA},'' \emph{IEEE Comm. Letters}, vol.~20, no.~10, pp.
  2055--2058, Oct. 2016.

\bibitem{N7}
Z.~Ding, M.~Peng, and H.~V. Poor, ``Cooperative non-orthogonal multiple access
  in {5G} systems,'' \emph{IEEE Comm. Letters}, vol.~19, no.~8, pp. 1462--1465,
  Aug. 2015.

\bibitem{N4}
Y.~Liu, Z.~Ding, M.~Elkashlan, and H.~V. Poor, ``Cooperative non-orthogonal
  multiple access with simultaneous wireless information and power transfer,''
  \emph{IEEE J. Select. Areas Commun.}, vol.~34, no.~4, pp. 938--953, Apr.
  2016.

\bibitem{N2}
Y.~Liu, Z.~Qin, M.~Elkashlan, Y.~Gao, and A.~Nallanathan, ``Non-orthogonal
  multiple access in massive {MIMO} aided heterogeneous networks,'' in
  \emph{Proc. of IEEE Global Communications Conference (GLOBECOM16)}, Dec.
  2016.

\bibitem{N6}
H.~Tabassum, E.~Hossain, and M.~J. Hossain, ``Modeling and analysis of uplink
  non-orthogonal multiple access in large-scale cellular networks using
  {P}oisson cluster processes,'' \emph{IEEE Trans. Commun.}, vol.~65, no.~8,
  pp. 3555--3570, Aug. 2017.

\bibitem{N5}
Y.~Liu, Z.~Ding, M.~Elkashlan, and J.~Yuan, ``Nonorthogonal multiple access in
  large-scale underlay cognitive radio networks,'' \emph{IEEE Trans. Vehicular
  Tech.}, vol.~65, no.~12, pp. 10\,152--10\,157, Dec. 2016.

\bibitem{N15}
J.~Zhu, J.~Wang, Y.~Huang, S.~He, X.~You, and L.~Yang, ``On optimal power
  allocation for downlink non-orthogonal multiple access systems,'' \emph{IEEE
  J. Selec. Areas Commun.}, vol.~35, no.~12, pp. 2744--2757, Dec. 2017.

\bibitem{N14}
C.~L. Wang, J.~Y. Chen, and Y.~J. Chen, ``Power allocation for a downlink
  non-orthogonal multiple access system,'' \emph{IEEE Wireless Comm. Letters},
  vol.~5, no.~5, pp. 532--535, Oct. 2016.

\bibitem{N15_19}
Y.~Sun, D.~W.~K. Ng, Z.~Ding, and R.~Schober, ``Optimal joint power and
  subcarrier allocation for full-duplex multicarrier non-orthogonal multiple
  access systems,'' \emph{IEEE Trans. Commun.}, vol.~65, no.~3, pp. 1077--1091,
  Mar. 2017.

\bibitem{my_nomaMag}
K.~S. Ali, H.~Elsawy, A.~Chaaban, and M.~S. Alouini, ``Non-orthogonal multiple
  access for large-scale {5G} networks: Interference aware design,'' \emph{IEEE
  Access}, vol.~5, pp. 21\,204--21\,216, 2017.

\bibitem{MH_Book2}
B.~Blaszczyszyn, M.~Haenggi, P.~Keeler, and S.~Mukherjee, \emph{Stochastic
  Geometry Analysis of Cellular Networks}.\hskip 1em plus 0.5em minus
  0.4em\relax Cambridge University Press, 2018.

\bibitem{3B1}
J.~Andrews, F.~Baccelli, and R.~Ganti, ``A tractable approach to coverage and
  rate in cellular networks,'' \emph{IEEE Trans. Commun.}, vol.~59, no.~11, pp.
  3122--3134, Nov. 2011.

\bibitem{h_tut}
H.~ElSawy, A.~Sultan-Salem, M.~S. Alouini, and M.~Z. Win, ``Modeling and
  analysis of cellular networks using stochastic geometry: A tutorial,''
  \emph{IEEE Commun. Surveys and Tutorials}, vol.~19, no.~1, pp. 167--203,
  2017.

\bibitem{di_renzo}
\BIBentryALTinterwordspacing
W.~Lu and M.~D. Renzo, ``Stochastic geometry modeling of cellular networks:
  Analysis, simulation and experimental validation,'' \emph{CoRR}, vol.
  abs/1506.03857, 2015. [Online]. Available:
  \url{http://arxiv.org/abs/1506.03857}
\BIBentrySTDinterwordspacing

\bibitem{N18}
Z.~Zhang, H.~Sun, and R.~Q. Hu, ``Downlink and uplink non-orthogonal multiple
  access in a dense wireless network,'' \emph{IEEE J. Selec. Areas Commun.},
  vol.~35, no.~12, pp. 2771--2784, Dec. 2017.

\bibitem{N16}
Z.~Zhang, H.~Sun, R.~Q. Hu, and Y.~Qian, ``Stochastic geometry based
  performance study on {5G} non-orthogonal multiple access scheme,'' in
  \emph{Proc. of IEEE Global Communications Conference (GLOBECOM16)}, Dec.
  2016, pp. 1--6.

\bibitem{comp1}
A.~H. Sakr and E.~Hossain, ``Location-aware cross-tier coordinated multipoint
  transmission in two-tier cellular networks,'' \emph{IEEE Trans. Wireless
  Commun.}, vol.~13, no.~11, pp. 6311--6325, Nov. 2014.

\bibitem{comp2}
A.~H. Sakr, H.~ElSawy, and E.~Hossain, ``Location-aware coordinated multipoint
  transmission in {OFDMA} networks,'' in \emph{Proc. of IEEE International
  Conference on Communications (ICC14)}, June 2014, pp. 5166--5171.

\bibitem{stienen3}
V.~Olsbo, ``On the correlation between the volumes of the typical {P}oisson
  {V}oronoi cell and the typical {S}tienen sphere,'' \emph{Advances in Applied
  Probability}, vol.~39, no.~4, pp. 883--892, 2007.

\bibitem{N6_20}
G.~Geraci, M.~Wildemeersch, and T.~Q.~S. Quek, ``Energy efficiency of
  distributed signal processing in wireless networks: A cross-layer analysis,''
  \emph{IEEE Trans. Signal Proc.}, vol.~64, no.~4, pp. 1034--1047, Feb. 2016.

\bibitem{N6_20_59}
M.~Wildemeersch, T.~Q.~S. Quek, M.~Kountouris, A.~Rabbachin, and C.~H. Slump,
  ``Successive interference cancellation in heterogeneous networks,''
  \emph{IEEE Trans. Commun.}, vol.~62, no.~12, pp. 4440--4453, Dec. 2014.

\bibitem{noma_sic}
H.~Sun, B.~Xie, R.~Q. Hu, and G.~Wu, ``Non-orthogonal multiple access with
  {SIC} error propagation in downlink wireless {MIMO} networks,'' in
  \emph{Proc. of IEEE 84th Vehicular Technology Conference (VTC Fall 2016)},
  Sep. 2016, pp. 1--5.

\bibitem{N16_18}
H.~A. David, \emph{Order statistics}.\hskip 1em plus 0.5em minus 0.4em\relax
  NJ: John Wiley, 1970.

\bibitem{N12}
J.~Choi, ``{NOMA}: Principles and recent results,'' in \emph{International
  Symposium on Wireless Communication Systems (ISWCS17)}, Aug. 2017, pp.
  349--354.

\end{thebibliography}

\end{document}